\documentclass[10pt,conference]{IEEEtran}

\usepackage{adjustbox}
\usepackage{algorithm}
\usepackage{algpseudocode}
\usepackage{amsmath}
\usepackage{amsfonts}
\usepackage[
  backend=biber,natbib=true,sorting=none,style=numeric-comp,maxbibnames=5
]{biblatex}
\usepackage{bookmark}
\usepackage{booktabs}
\usepackage[skip=0.3\baselineskip]{caption}
\usepackage{colortbl}
\usepackage[inline]{enumitem}
\usepackage{graphicx}
\usepackage{hyperref}
\usepackage{listings}
\usepackage{makecell}
\usepackage{mathtools}
\usepackage{multicol}
\usepackage{multirow}
\usepackage{nameref}
\usepackage{namedef}
\usepackage{pbox}
\usepackage{pgffor}
\usepackage{soul}
\usepackage{subcaption}
\usepackage{tabularx}
\usepackage{tcolorbox}
\usepackage{textcomp}
\usepackage{varioref}
\usepackage{xcolor}
\usepackage{xspace}
\usepackage{xstring}
\usepackage{tablefootnote}
\usepackage{fontawesome5}
\usepackage{sparklines}

\let\oldttfamily\ttfamily
\renewcommand{\ttfamily}{\oldttfamily\small}

\definecolor{backcolour}{rgb}{0.95,0.95,0.92}

\newcolumntype{Y}{>{\centering\arraybackslash}X}
\newcolumntype{Z}{>{\rightline\arraybackslash}X}

\newif\ifshowcomments
\showcommentstrue 

\newcommand{\red}[1]{\textcolor{red}{#1}}

\newcommand{\hide}[1]{\ignorespaces}

\newcommand{\migbench}{\textsc{PyMigBench}\xspace}

\newcommand{\taxonomy}{\textsc{PyMigTax}\xspace}

\newcommand{\etal}{et al.\xspace}
\newcommand{\eg}{e.g.,\xspace}
\newcommand{\ie}{i.e.\xspace}

\newcommand{\rqBench}[0]{\ref{rq:benchmark}\xspace}
\newcommand{\rqTest}[0]{\ref{rq:test}\xspace}
\newcommand{\rqUnseen}[0]{\ref{rq:unseen}\xspace}

\newcommand{\ccs}[0]{code changes\xspace}
\newcommand{\cc}[0]{code change\xspace}

\newcommand{\codechange}[0]{migration-related code change\xspace}

\newcommand{\codechanges}[0]{migration-related code changes\xspace}

\newcommand{\pre}[1]{#1$_{pre}$\xspace}
\newcommand{\dev}[1]{#1$_{dev}$\xspace}
\newcommand{\llm}[1]{#1$_{llm}$\xspace}

\newcommand{\codePre}{\pre{code}}
\newcommand{\codeDev}{\dev{code}}
\newcommand{\codeLLM}{\llm{code}}
\newcommand{\filePre}{\pre{File}}
\newcommand{\fileDev}{\dev{File}}
\newcommand{\fileLLM}{\llm{File}}

\newcommand{\changeDev}{\dev{Change}}
\newcommand{\changeLLM}{\llm{Change}}

\newcommand{\uChangeDev}{\dev{uChange}}
\newcommand{\uChangeLLM}{\llm{uChange}}

\newcommand{\pair}[2]{\lib{#1}{\textrightarrow}\lib{#2}}

\newcommand{\MigStatusCorrect}{Correct\xspace}
\newcommand{\MigStatusExtraChange}{Correct with non-refactoring changes\xspace}

\newcommand{\MigStatusPartialCorrect}{Partially correct\xspace}
\newcommand{\MigStatusIncorrect}{Incorrect\xspace}
\newcommand{\MigStatusSyntaxError}{Syntax error\xspace}
\newcommand{\MigStatusResponseFailure}{Response failure\xspace}
\newcommand{\MigStatusResponseFailureLower}{response failure\xspace}

\newcommand{\CCStatusCorrect}{Correct\xspace}

\newcommand{\CCStatusIncorrect}{Incorrect\xspace}

\newcommand{\fcall}[0]{{function call}\xspace}
\newcommand{\Fcall}[0]{{Function call}\xspace}
\newcommand{\fref}[0]{{function reference}\xspace}

\newcommand{\Type}[0]{{Type}\xspace}
\newcommand{\type}[0]{{type}\xspace}
\newcommand{\imp}[0]{import\xspace}

\newcommand{\Imp}[0]{{Import}\xspace}
\newcommand{\exc}[0]{{exception}\xspace}
\newcommand{\Exc}[0]{{Exception}\xspace}
\newcommand{\attr}[0]{{attribute}\xspace}
\newcommand{\Attr}[0]{{Attribute}\xspace}
\newcommand{\dec}[0]{{decorator}\xspace}
\newcommand{\Dec}[0]{{Decorator}\xspace}

\newcommand{\oo}[0]{one-to-one\xspace}
\newcommand{\OO}[0]{One-to-One\xspace}
\newcommand{\om}[0]{one-to-many\xspace}
\newcommand{\OM}[0]{One-to-Many\xspace}
\newcommand{\mo}[0]{many-to-one\xspace}
\newcommand{\MO}[0]{Many-to-One\xspace}
\newcommand{\mm}[0]{many-to-many\xspace}
\newcommand{\MM}[0]{Many-to-Many\xspace}

\newcommand{\ZO}[0]{Zero-to-One\xspace}

\newcommand{\OZ}[0]{One-to-Zero\xspace}

\newcommand{\argAdd}[0]{{argument addition}\xspace}

\newcommand{\argDel}[0]{{argument deletion}\xspace}

\newcommand{\argTrans}[0]{{argument transformation}\xspace}
\newcommand{\ArgTrans}[0]{{Argument transformation}\xspace}
\newcommand{\enc}[0]{{element name change}\xspace}

\newcommand{\ElemNC}[0]{{Element name change}\xspace}

\newcommand{\noProps}[0]{{no properties}\xspace}
\newcommand{\NoProps}[0]{{No properties}\xspace}

\newcommand{\argNC}[0]{{argument name change}\xspace}
\newcommand{\ArgNC}[0]{{Argument name change}\xspace}
\newcommand{\paramAdd}[0]{{parameter addition to decorated function}\xspace}

\newcommand{\lib}[1]{\textit{#1}\xspace}

\definecolor{rem}{RGB}{255,40,80}
\definecolor{add}{RGB}{20,177,64}

\newcommand{\segment}[1]{\texttt{#1}}

\newcommand{\mini}[0]{GPT-4o mini\xspace}
\newcommand{\fouro}[0]{GPT-4o\xspace}
\newcommand{\llama}[0]{LLama 3.1\xspace}

\newcommand{\mn}[0]{Mini\xspace}
\newcommand{\fo}[0]{4o\xspace}
\newcommand{\lm}[0]{LLama\xspace}

\newcommand{\paramAddCell}{{Param addition}}
\newcommand{\argTransCell}{{Arg transform}}
\newcommand{\argAddCell}{Arg addition}
\newcommand{\argDelCell}{Arg deletion}
\newcommand{\argNCCell}{{Arg name change}}
\newcommand{\asyncTransCell}{{Async transform}}
\newcommand{\outTransCell}{Output transform}
\newcommand{\elemNCCell}{Elmnt name change}
\newcommand{\fRefCell}{Function reference}

\newcommand{\evalCountCell}{\#Eval}
\newcommand{\correctPercentCell}{\%Cor}

\usepackage[absolute,overlay]{textpos}
\usepackage{tikz}
\usepackage{xcolor}
\usepackage{calc}

\renewcommand{\arraystretch}{1.2} 

\newcounter{findingctr}

\newcommand{\finding}[2]{\refstepcounter{findingctr}\emph{#1:\label{box:#2}}}

\newlength{\boxw}
\newlength{\boxh}
\newlength{\shadowsize}
\newlength{\boxroundness}
\newlength{\tmpa}
\newsavebox{\shadowblockbox}

\newenvironment{findingenv}[2]%
{\vspace{0.2cm}\noindent
\begin{lrbox}{
\shadowblockbox
}
\begin{minipage}{.98\columnwidth}
\finding{#1}{#2}~}%
{\end{minipage}\end{lrbox}%
\settowidth{\boxw}{\usebox{\shadowblockbox}}%
\settodepth{\tmpa}{\usebox{\shadowblockbox}}%
\settoheight{\boxh}{\usebox{\shadowblockbox}}%
\addtolength{\boxh}{\tmpa}%
\begin{tikzpicture}
\addtolength{\boxw}{\boxroundness * 2}
\addtolength{\boxh}{\boxroundness * 2}

\foreach \x in {0,.05,...,1}
{
\setlength{\tmpa}{\shadowsize * \real{\x}}
\fill[xshift=\shadowsize - 1pt,yshift=-\shadowsize + 
1pt,black,opacity=.04,rounded corners=\boxroundness] 
(\tmpa, \tmpa) rectangle +(\boxw - \tmpa - \tmpa, \boxh - \tmpa - 
\tmpa);
}

\filldraw[fill=white!50, draw=black!80, rounded corners=\boxroundness] (0, 
0) rectangle (\boxw, \boxh);
\draw node[xshift=\boxroundness,yshift=\boxroundness,inner sep=0pt,outer 
sep=0pt,anchor=south west] (0,0) {\usebox{\shadowblockbox}};
\end{tikzpicture}\vspace{0cm}%
}

\newcommand{\ArgAddTotalCount}{578\xspace}
\newcommand{\ArgAddTotalPercent}{19\%\xspace}

\newcommand{\ArgDelTotalCount}{349\xspace}
\newcommand{\ArgDelTotalPercent}{12\%\xspace}

\newcommand{\ArgNCTotalCount}{112\xspace}
\newcommand{\ArgNCTotalPercent}{3.7\%\xspace}

\newcommand{\ArgTransTotalCount}{317\xspace}
\newcommand{\ArgTransTotalPercent}{11\%\xspace}

\newcommand{\AsyncTransTotalCount}{50\xspace}
\newcommand{\AsyncTransTotalPercent}{1.7\%\xspace}

\newcommand{\AttrTotalCount}{183\xspace}
\newcommand{\AttrTotalPercent}{6.1\%\xspace}
\newcommand{\CCAllCount}{2,989\xspace}

\newcommand{\CCArgparseClickCorrectPercent}{94\%\xspace}

\newcommand{\CCExperimentCount}{2,910\xspace}

\newcommand{\DecTotalCount}{411\xspace}
\newcommand{\DecTotalPercent}{14\%\xspace}
\newcommand{\DomainAllCount}{34\xspace}

\newcommand{\ElemNCTotalCount}{1,025\xspace}
\newcommand{\ElemNCTotalPercent}{34\%\xspace}

\newcommand{\ExcTotalCount}{29\xspace}
\newcommand{\ExcTotalPercent}{1.0\%\xspace}

\newcommand{\FCallTotalCount}{1,804\xspace}
\newcommand{\FCallTotalPercent}{60\%\xspace}

\newcommand{\FRefTotalCount}{12\xspace}
\newcommand{\FRefTotalPercent}{0.4\%\xspace}

\newcommand{\FouroAvgDiff}{4.8\%\xspace}

\newcommand{\FouroCCBMArgAddCorrectPercent}{88\%\xspace}
\newcommand{\FouroCCBMArgAddTotalCount}{244\xspace}

\newcommand{\FouroCCBMArgDelCorrectPercent}{89\%\xspace}
\newcommand{\FouroCCBMArgDelTotalCount}{234\xspace}

\newcommand{\FouroCCBMArgNCCorrectPercent}{84\%\xspace}
\newcommand{\FouroCCBMArgNCTotalCount}{58\xspace}

\newcommand{\FouroCCBMArgTransCorrectPercent}{77\%\xspace}
\newcommand{\FouroCCBMArgTransTotalCount}{220\xspace}

\newcommand{\FouroCCBMAsyncTransCorrectPercent}{92\%\xspace}
\newcommand{\FouroCCBMAsyncTransTotalCount}{50\xspace}

\newcommand{\FouroCCBMAttrCorrectPercent}{89\%\xspace}
\newcommand{\FouroCCBMAttrTotalCount}{103\xspace}

\newcommand{\FouroCCBMDecCorrectPercent}{87\%\xspace}
\newcommand{\FouroCCBMDecTotalCount}{353\xspace}

\newcommand{\FouroCCBMElemNCCorrectPercent}{90\%\xspace}
\newcommand{\FouroCCBMElemNCTotalCount}{771\xspace}

\newcommand{\FouroCCBMEvaluatedCount}{1,996\xspace}

\newcommand{\FouroCCBMExcCorrectPercent}{93\%\xspace}
\newcommand{\FouroCCBMExcTotalCount}{29\xspace}

\newcommand{\FouroCCBMExcludedCount}{914\xspace}

\newcommand{\FouroCCBMFCallCorrectPercent}{93\%\xspace}
\newcommand{\FouroCCBMFCallTotalCount}{1,092\xspace}

\newcommand{\FouroCCBMFRefCorrectPercent}{100\%\xspace}
\newcommand{\FouroCCBMFRefTotalCount}{12\xspace}

\newcommand{\FouroCCBMHigherCardinalityCorrectPercent}{84\%\xspace}
\newcommand{\FouroCCBMHigherCardinalityTotalCount}{164\xspace}

\newcommand{\FouroCCBMImpCorrectPercent}{98\%\xspace}
\newcommand{\FouroCCBMImpTotalCount}{593\xspace}

\newcommand{\FouroCCBMMMCorrectPercent}{89\%\xspace}
\newcommand{\FouroCCBMMMTotalCount}{27\xspace}

\newcommand{\FouroCCBMMOCorrectPercent}{91\%\xspace}
\newcommand{\FouroCCBMMOTotalCount}{34\xspace}

\newcommand{\FouroCCBMNoPropsCorrectPercent}{97\%\xspace}
\newcommand{\FouroCCBMNoPropsTotalCount}{150\xspace}

\newcommand{\FouroCCBMOMCorrectPercent}{80\%\xspace}
\newcommand{\FouroCCBMOMTotalCount}{103\xspace}

\newcommand{\FouroCCBMOOCorrectPercent}{91\%\xspace}
\newcommand{\FouroCCBMOOTotalCount}{894\xspace}

\newcommand{\FouroCCBMOZCorrectPercent}{100\%\xspace}
\newcommand{\FouroCCBMOZTotalCount}{290\xspace}

\newcommand{\FouroCCBMOutTransCorrectPercent}{92\%\xspace}
\newcommand{\FouroCCBMOutTransTotalCount}{49\xspace}

\newcommand{\FouroCCBMParamAddCorrectPercent}{90\%\xspace}
\newcommand{\FouroCCBMParamAddTotalCount}{127\xspace}

\newcommand{\FouroCCBMTotalCorrectCount}{1,867\xspace}
\newcommand{\FouroCCBMTotalCorrectPercent}{94\%\xspace}
\newcommand{\FouroCCBMTotalIncorrectCount}{129\xspace}
\newcommand{\FouroCCBMTotalIncorrectPercent}{6.5\%\xspace}

\newcommand{\FouroCCBMTypeCorrectPercent}{88\%\xspace}
\newcommand{\FouroCCBMTypeTotalCount}{34\xspace}

\newcommand{\FouroCCBMZOCorrectPercent}{89\%\xspace}
\newcommand{\FouroCCBMZOTotalCount}{55\xspace}

\newcommand{\FouroCostPerKLOC}{0.16\xspace}
\newcommand{\FouroMigBMAtLeastOneCorrectCount}{296\xspace}
\newcommand{\FouroMigBMAtLeastOneCorrectPercent}{94\%\xspace}
\newcommand{\FouroMigBMCorrectCount}{179\xspace}
\newcommand{\FouroMigBMCorrectPercent}{57\%\xspace}
\newcommand{\FouroMigBMCorrectWithExtraChangesCount}{55\xspace}
\newcommand{\FouroMigBMCorrectWithExtraChangesPercent}{18\%\xspace}

\newcommand{\FouroMigBMIncorrectCount}{2\xspace}
\newcommand{\FouroMigBMIncorrectPercent}{0.6\%\xspace}
\newcommand{\FouroMigBMNoCorrectCount}{18\xspace}
\newcommand{\FouroMigBMNoCorrectPercent}{5.7\%\xspace}
\newcommand{\FouroMigBMPartiallyCorrectCount}{62\xspace}
\newcommand{\FouroMigBMPartiallyCorrectPercent}{20\%\xspace}
\newcommand{\FouroMigBMResponseFailureCount}{13\xspace}
\newcommand{\FouroMigBMResponseFailurePercent}{4.1\%\xspace}
\newcommand{\FouroMigBMSyntaxErrorCount}{3\xspace}
\newcommand{\FouroMigBMSyntaxErrorPercent}{1.0\%\xspace}
\newcommand{\FouroMigTestCorrectCount}{16\xspace}
\newcommand{\FouroMigTestCorrectPercent}{64\%\xspace}
\newcommand{\FouroMigTestIncorrectCount}{7\xspace}
\newcommand{\FouroMigTestIncorrectPercent}{28\%\xspace}
\newcommand{\FouroMigTestPartiallyCorrectCount}{2\xspace}
\newcommand{\FouroMigTestPartiallyCorrectPercent}{8.0\%\xspace}

\newcommand{\FouroTotalCost}{420\xspace}

\newcommand{\HigherCardinalityLibPairPercent}{34\%\xspace}

\newcommand{\HigherCardinalityTotalCount}{210\xspace}
\newcommand{\HigherCardinalityTotalPercent}{7.0\%\xspace}

\newcommand{\ImpTotalCount}{732\xspace}
\newcommand{\ImpTotalPercent}{24\%\xspace}

\newcommand{\KappaOne}{0.73\xspace}
\newcommand{\KappaTwo}{0.83\xspace}

\newcommand{\LibPairAllCount}{134\xspace}
\newcommand{\LlamaAvgDiff}{6.6\%\xspace}

\newcommand{\LlamaCCBMArgAddCorrectPercent}{85\%\xspace}
\newcommand{\LlamaCCBMArgAddTotalCount}{150\xspace}

\newcommand{\LlamaCCBMArgDelCorrectPercent}{85\%\xspace}
\newcommand{\LlamaCCBMArgDelTotalCount}{175\xspace}

\newcommand{\LlamaCCBMArgNCCorrectPercent}{60\%\xspace}
\newcommand{\LlamaCCBMArgNCTotalCount}{20\xspace}

\newcommand{\LlamaCCBMArgTransCorrectPercent}{77\%\xspace}
\newcommand{\LlamaCCBMArgTransTotalCount}{115\xspace}

\newcommand{\LlamaCCBMAsyncTransCorrectPercent}{83\%\xspace}
\newcommand{\LlamaCCBMAsyncTransTotalCount}{35\xspace}

\newcommand{\LlamaCCBMAttrCorrectPercent}{83\%\xspace}
\newcommand{\LlamaCCBMAttrTotalCount}{63\xspace}

\newcommand{\LlamaCCBMDecCorrectPercent}{87\%\xspace}
\newcommand{\LlamaCCBMDecTotalCount}{197\xspace}

\newcommand{\LlamaCCBMElemNCCorrectPercent}{84\%\xspace}
\newcommand{\LlamaCCBMElemNCTotalCount}{398\xspace}

\newcommand{\LlamaCCBMEvaluatedCount}{911\xspace}

\newcommand{\LlamaCCBMExcCorrectPercent}{65\%\xspace}
\newcommand{\LlamaCCBMExcTotalCount}{17\xspace}

\newcommand{\LlamaCCBMExcludedCount}{1,999\xspace}

\newcommand{\LlamaCCBMFCallCorrectPercent}{88\%\xspace}
\newcommand{\LlamaCCBMFCallTotalCount}{518\xspace}

\newcommand{\LlamaCCBMFRefCorrectPercent}{33\%\xspace}
\newcommand{\LlamaCCBMFRefTotalCount}{3\xspace}

\newcommand{\LlamaCCBMHigherCardinalityCorrectPercent}{70\%\xspace}
\newcommand{\LlamaCCBMHigherCardinalityTotalCount}{94\xspace}

\newcommand{\LlamaCCBMImpCorrectPercent}{95\%\xspace}
\newcommand{\LlamaCCBMImpTotalCount}{276\xspace}

\newcommand{\LlamaCCBMMMCorrectPercent}{82\%\xspace}
\newcommand{\LlamaCCBMMMTotalCount}{17\xspace}

\newcommand{\LlamaCCBMMOCorrectPercent}{62\%\xspace}
\newcommand{\LlamaCCBMMOTotalCount}{16\xspace}

\newcommand{\LlamaCCBMNoPropsCorrectPercent}{84\%\xspace}
\newcommand{\LlamaCCBMNoPropsTotalCount}{43\xspace}

\newcommand{\LlamaCCBMOMCorrectPercent}{69\%\xspace}
\newcommand{\LlamaCCBMOMTotalCount}{61\xspace}

\newcommand{\LlamaCCBMOOCorrectPercent}{86\%\xspace}
\newcommand{\LlamaCCBMOOTotalCount}{408\xspace}

\newcommand{\LlamaCCBMOZCorrectPercent}{99\%\xspace}
\newcommand{\LlamaCCBMOZTotalCount}{99\xspace}

\newcommand{\LlamaCCBMOutTransCorrectPercent}{88\%\xspace}
\newcommand{\LlamaCCBMOutTransTotalCount}{25\xspace}

\newcommand{\LlamaCCBMParamAddCorrectPercent}{87\%\xspace}
\newcommand{\LlamaCCBMParamAddTotalCount}{110\xspace}

\newcommand{\LlamaCCBMTotalCorrectCount}{808\xspace}
\newcommand{\LlamaCCBMTotalCorrectPercent}{89\%\xspace}
\newcommand{\LlamaCCBMTotalIncorrectCount}{103\xspace}
\newcommand{\LlamaCCBMTotalIncorrectPercent}{11\%\xspace}

\newcommand{\LlamaCCBMTypeCorrectPercent}{75\%\xspace}
\newcommand{\LlamaCCBMTypeTotalCount}{12\xspace}

\newcommand{\LlamaCCBMZOCorrectPercent}{97\%\xspace}
\newcommand{\LlamaCCBMZOTotalCount}{34\xspace}

\newcommand{\LlamaMigBMAtLeastOneCorrectCount}{159\xspace}
\newcommand{\LlamaMigBMAtLeastOneCorrectPercent}{51\%\xspace}
\newcommand{\LlamaMigBMCorrectCount}{83\xspace}
\newcommand{\LlamaMigBMCorrectPercent}{26\%\xspace}
\newcommand{\LlamaMigBMCorrectWithExtraChangesCount}{28\xspace}
\newcommand{\LlamaMigBMCorrectWithExtraChangesPercent}{8.9\%\xspace}

\newcommand{\LlamaMigBMIncorrectCount}{7\xspace}
\newcommand{\LlamaMigBMIncorrectPercent}{2.2\%\xspace}
\newcommand{\LlamaMigBMNoCorrectCount}{155\xspace}
\newcommand{\LlamaMigBMNoCorrectPercent}{49\%\xspace}
\newcommand{\LlamaMigBMPartiallyCorrectCount}{48\xspace}
\newcommand{\LlamaMigBMPartiallyCorrectPercent}{15\%\xspace}
\newcommand{\LlamaMigBMResponseFailureCount}{132\xspace}
\newcommand{\LlamaMigBMResponseFailurePercent}{42\%\xspace}
\newcommand{\LlamaMigBMSyntaxErrorCount}{16\xspace}
\newcommand{\LlamaMigBMSyntaxErrorPercent}{5.1\%\xspace}
\newcommand{\LlamaMigTestCorrectCount}{9\xspace}
\newcommand{\LlamaMigTestCorrectPercent}{36\%\xspace}
\newcommand{\LlamaMigTestIncorrectCount}{15\xspace}
\newcommand{\LlamaMigTestIncorrectPercent}{60\%\xspace}
\newcommand{\LlamaMigTestPartiallyCorrectCount}{1\xspace}
\newcommand{\LlamaMigTestPartiallyCorrectPercent}{4.0\%\xspace}

\newcommand{\MMTotalCount}{27\xspace}
\newcommand{\MMTotalPercent}{0.9\%\xspace}

\newcommand{\MOTotalCount}{41\xspace}
\newcommand{\MOTotalPercent}{1.4\%\xspace}
\newcommand{\MigAllCount}{321\xspace}

\newcommand{\MigCloneFailedCount}{2\xspace}
\newcommand{\MigCloneSuccessCount}{319\xspace}
\newcommand{\MigCoveredCount}{27\xspace}
\newcommand{\MigCoveredPassingCount}{25\xspace}
\newcommand{\MigErrorRunningTestCount}{172\xspace}
\newcommand{\MigExperimentCount}{314\xspace}
\newcommand{\MigHasTestsCount}{218\xspace}

\newcommand{\MigSyntaxErrorCount}{5\xspace}
\newcommand{\MigTestEvaluatedCount}{25\xspace}

\newcommand{\MigTestExecutedCount}{46\xspace}
\newcommand{\MiniAvgDiff}{4.0\%\xspace}

\newcommand{\MiniCCBMArgAddCorrectPercent}{80\%\xspace}
\newcommand{\MiniCCBMArgAddTotalCount}{244\xspace}

\newcommand{\MiniCCBMArgDelCorrectPercent}{87\%\xspace}
\newcommand{\MiniCCBMArgDelTotalCount}{234\xspace}

\newcommand{\MiniCCBMArgNCCorrectPercent}{59\%\xspace}
\newcommand{\MiniCCBMArgNCTotalCount}{58\xspace}

\newcommand{\MiniCCBMArgTransCorrectPercent}{76\%\xspace}
\newcommand{\MiniCCBMArgTransTotalCount}{201\xspace}

\newcommand{\MiniCCBMAsyncTransCorrectPercent}{90\%\xspace}
\newcommand{\MiniCCBMAsyncTransTotalCount}{50\xspace}

\newcommand{\MiniCCBMAttrCorrectPercent}{78\%\xspace}
\newcommand{\MiniCCBMAttrTotalCount}{103\xspace}

\newcommand{\MiniCCBMDecCorrectPercent}{90\%\xspace}
\newcommand{\MiniCCBMDecTotalCount}{335\xspace}

\newcommand{\MiniCCBMElemNCCorrectPercent}{85\%\xspace}
\newcommand{\MiniCCBMElemNCTotalCount}{773\xspace}

\newcommand{\MiniCCBMEvaluatedCount}{1,966\xspace}

\newcommand{\MiniCCBMExcCorrectPercent}{86\%\xspace}
\newcommand{\MiniCCBMExcTotalCount}{29\xspace}

\newcommand{\MiniCCBMExcludedCount}{944\xspace}

\newcommand{\MiniCCBMFCallCorrectPercent}{87\%\xspace}
\newcommand{\MiniCCBMFCallTotalCount}{1,093\xspace}

\newcommand{\MiniCCBMFRefCorrectPercent}{58\%\xspace}
\newcommand{\MiniCCBMFRefTotalCount}{12\xspace}

\newcommand{\MiniCCBMHigherCardinalityCorrectPercent}{79\%\xspace}
\newcommand{\MiniCCBMHigherCardinalityTotalCount}{164\xspace}

\newcommand{\MiniCCBMImpCorrectPercent}{94\%\xspace}
\newcommand{\MiniCCBMImpTotalCount}{580\xspace}

\newcommand{\MiniCCBMMMCorrectPercent}{78\%\xspace}
\newcommand{\MiniCCBMMMTotalCount}{27\xspace}

\newcommand{\MiniCCBMMOCorrectPercent}{82\%\xspace}
\newcommand{\MiniCCBMMOTotalCount}{34\xspace}

\newcommand{\MiniCCBMNoPropsCorrectPercent}{88\%\xspace}
\newcommand{\MiniCCBMNoPropsTotalCount}{150\xspace}

\newcommand{\MiniCCBMOMCorrectPercent}{78\%\xspace}
\newcommand{\MiniCCBMOMTotalCount}{103\xspace}

\newcommand{\MiniCCBMOOCorrectPercent}{84\%\xspace}
\newcommand{\MiniCCBMOOTotalCount}{877\xspace}

\newcommand{\MiniCCBMOZCorrectPercent}{99\%\xspace}
\newcommand{\MiniCCBMOZTotalCount}{290\xspace}

\newcommand{\MiniCCBMOutTransCorrectPercent}{78\%\xspace}
\newcommand{\MiniCCBMOutTransTotalCount}{49\xspace}

\newcommand{\MiniCCBMParamAddCorrectPercent}{95\%\xspace}
\newcommand{\MiniCCBMParamAddTotalCount}{127\xspace}

\newcommand{\MiniCCBMTotalCorrectCount}{1,744\xspace}
\newcommand{\MiniCCBMTotalCorrectPercent}{89\%\xspace}
\newcommand{\MiniCCBMTotalIncorrectCount}{222\xspace}
\newcommand{\MiniCCBMTotalIncorrectPercent}{11\%\xspace}

\newcommand{\MiniCCBMTypeCorrectPercent}{85\%\xspace}
\newcommand{\MiniCCBMTypeTotalCount}{34\xspace}

\newcommand{\MiniCCBMZOCorrectPercent}{87\%\xspace}
\newcommand{\MiniCCBMZOTotalCount}{55\xspace}

\newcommand{\MiniCostPerKLOC}{0.01\xspace}
\newcommand{\MiniMigBMAtLeastOneCorrectCount}{291\xspace}
\newcommand{\MiniMigBMAtLeastOneCorrectPercent}{93\%\xspace}
\newcommand{\MiniMigBMCorrectCount}{154\xspace}
\newcommand{\MiniMigBMCorrectPercent}{49\%\xspace}
\newcommand{\MiniMigBMCorrectWithExtraChangesCount}{53\xspace}
\newcommand{\MiniMigBMCorrectWithExtraChangesPercent}{17\%\xspace}

\newcommand{\MiniMigBMIncorrectCount}{6\xspace}
\newcommand{\MiniMigBMIncorrectPercent}{1.9\%\xspace}
\newcommand{\MiniMigBMNoCorrectCount}{23\xspace}
\newcommand{\MiniMigBMNoCorrectPercent}{7.3\%\xspace}
\newcommand{\MiniMigBMPartiallyCorrectCount}{84\xspace}
\newcommand{\MiniMigBMPartiallyCorrectPercent}{27\%\xspace}
\newcommand{\MiniMigBMResponseFailureCount}{11\xspace}
\newcommand{\MiniMigBMResponseFailurePercent}{3.5\%\xspace}
\newcommand{\MiniMigBMSyntaxErrorCount}{6\xspace}
\newcommand{\MiniMigBMSyntaxErrorPercent}{1.9\%\xspace}
\newcommand{\MiniMigTestCorrectCount}{13\xspace}
\newcommand{\MiniMigTestCorrectPercent}{52\%\xspace}
\newcommand{\MiniMigTestIncorrectCount}{7\xspace}
\newcommand{\MiniMigTestIncorrectPercent}{28\%\xspace}
\newcommand{\MiniMigTestPartiallyCorrectCount}{5\xspace}
\newcommand{\MiniMigTestPartiallyCorrectPercent}{20\%\xspace}

\newcommand{\MiniTotalCost}{15\xspace}

\newcommand{\NoPropsTotalCount}{262\xspace}
\newcommand{\NoPropsTotalPercent}{8.8\%\xspace}

\newcommand{\OMTotalCount}{142\xspace}
\newcommand{\OMTotalPercent}{4.8\%\xspace}

\newcommand{\OOTotalCount}{1,552\xspace}
\newcommand{\OOTotalPercent}{52\%\xspace}

\newcommand{\OZTotalCount}{297\xspace}
\newcommand{\OZTotalPercent}{10\%\xspace}

\newcommand{\OutTransTotalCount}{49\xspace}
\newcommand{\OutTransTotalPercent}{1.6\%\xspace}

\newcommand{\ParamAddTotalCount}{127\xspace}
\newcommand{\ParamAddTotalPercent}{4.2\%\xspace}

\newcommand{\RepoExperimentCount}{294\xspace}

\newcommand{\TypeTotalCount}{55\xspace}
\newcommand{\TypeTotalPercent}{1.8\%\xspace}

\newcommand{\ZOTotalCount}{198\xspace}
\newcommand{\ZOTotalPercent}{6.6\%\xspace}

\newcommand{\UnseenFouroAllPass}{5\xspace}
\newcommand{\UnseenFouroAllPassPercent}{50\%\xspace}

\newcommand{\UnseenFouroMedianPassedTests}{100\%\xspace}

\newcommand{\UnseenFouroNoPass}{2\xspace}

\newcommand{\UnseenLibPairsCount}{10\xspace}
\newcommand{\UnseenLlamaAllPass}{1\xspace}
\newcommand{\UnseenLlamaAllPassPercent}{10\%\xspace}

\newcommand{\UnseenLlamaMedianPassedTests}{0.0\%\xspace}

\newcommand{\UnseenLlamaNoPassPercent}{70\%\xspace}

\newcommand{\UnseenMaxApisCount}{210\xspace}
\newcommand{\UnseenMaxFilesCount}{5\xspace}
\newcommand{\UnseenMaxModifiedLines}{1,125\xspace}
\newcommand{\UnseenMaxTestsCount}{903\xspace}
\newcommand{\UnseenMedianApisCount}{6\xspace}
\newcommand{\UnseenMedianFilesCount}{1\xspace}
\newcommand{\UnseenMedianModifiedLines}{36\xspace}
\newcommand{\UnseenMedianTestsCount}{83\xspace}
\newcommand{\UnseenMigsCount}{10\xspace}
\newcommand{\UnseenMinApisCount}{2\xspace}
\newcommand{\UnseenMinFilesCount}{1\xspace}
\newcommand{\UnseenMinModifiedLines}{3\xspace}
\newcommand{\UnseenMinTestsCount}{7\xspace}
\newcommand{\UnseenMiniAllPass}{4\xspace}
\newcommand{\UnseenMiniAllPassPercent}{40\%\xspace}

\newcommand{\UnseenMiniMedianPassedTests}{47\%\xspace}

\newcommand{\UnseenMiniNoPass}{4\xspace}

\newcommand{\UnseenNewLibPairsCount}{6\xspace}
\newcommand{\UnseenOldLibPairsCount}{4\xspace}
\newcommand{\UnseenReposCount}{10\xspace}

\newcommand{\Description}[1]{} 

\usepackage{fancyhdr}

\begin{document}

    \title{An Empirical Study of Python Library Migration Using Large Language Models}
    \pagestyle{plain}

\def\emptyBlockWidth{20pt}


\author{  

  \IEEEauthorblockN{\makebox[.33\textwidth][c]{Mohayeminul Islam}}
  \IEEEauthorblockA{\textit{University of Alberta} \\    
   Canada \\
   mohayemin@ualberta.ca}
 
 \and
 
  \IEEEauthorblockN{\makebox[.33\textwidth][c]{Ajay Kumar Jha}}
  \IEEEauthorblockA{\textit{North Dakota State University} \\
   USA \\
   ajay.jha.1@ndsu.edu}
 
   \and
 
   \IEEEauthorblockN{\makebox[.33\textwidth][c]{May Mahmoud}}
   \IEEEauthorblockA{\textit{New York University Abu Dhabi} \\
   United Arab Emirates \\
   m.mahmoud@nyu.edu}   

  \and 

  \IEEEauthorblockN{\makebox[.17\textwidth][c]{~}} 

  \and

   \IEEEauthorblockN{\makebox[.33\textwidth][c]{Ildar Akhmetov}} 
   \IEEEauthorblockA{\textit{Northeastern University} \\
   Canada \\
   i.akhmetov@northeastern.edu}
 
   \and
 
   \IEEEauthorblockN{\makebox[.33\textwidth][c]{Sarah Nadi}} 
   \IEEEauthorblockA{\textit{New York University Abu Dhabi} \\
   United Arab Emirates \\
   sarah.nadi@nyu.edu}
  
   \and

   \IEEEauthorblockN{\makebox[.17\textwidth][c]{~}} 
}

    \maketitle

    \pagestyle{fancy}

    \fancyfoot[R]{\red{preprint}}

    \begin{abstract}
Library migration is the process of replacing one library with another library that provides similar functionality.
Manual library migration is time consuming and error prone,
as it requires developers to understand the APIs of both libraries, map them, and perform the necessary code transformations.
Large Language Models (LLMs) are shown to be effective at generating and transforming code as well as finding similar code, which are necessary upstream tasks for library migration. 
Such capabilities suggest that LLMs may be suitable for library migration.
Accordingly, this paper investigates the effectiveness of LLMs for migration between Python libraries. 
We evaluate three LLMs, \llama, \mini, and \fouro on \migbench, where we migrate \MigAllCount real-world library migrations that include \CCAllCount \codechanges.
To measure correctness, we (1) compare the LLM's migrated code with the developers' migrated code in the benchmark and (2) run the unit tests available in the client repositories. 
We find that \llama, \mini, and \fouro correctly migrate \LlamaCCBMTotalCorrectPercent, \MiniCCBMTotalCorrectPercent, and \FouroCCBMTotalCorrectPercent of the \codechanges, respectively.
We also find that \LlamaMigTestCorrectPercent, \MiniMigTestCorrectPercent and \FouroMigTestCorrectPercent of the \llama, \mini, and \fouro migrations pass the same tests that passed in the developer's migration.
To ensure the LLMs are not reciting the migrations, we also evaluate them on 10 new repositories where the migration never happened.
Overall, our results suggest that LLMs can be effective in migrating code between libraries, but we also identify some open challenges.
    \end{abstract}

    \section{Introduction}
\label{sec:intro}
Software libraries are essential for modern software development, as they provide reusable code that can significantly reduce development time and effort.
Developers often replace one library with another in their applications to improve performance, address security vulnerabilities, or even for license compatibility \cite{keepMeUpdated, kula2018developers}.
This process, known as \textit{library migration} \cite{teyton2012mining}, is time consuming and error prone \cite{kula2018developers},
as it requires developers to understand the Application Programming Interfaces (APIs) of both libraries, find API replacements (\textit{API mapping}), and perform various code transformations~\cite{alrubaye2019use}.

There have been previous efforts towards automating library migration, but fully usable tools for this task are limited.
With the exception of a few attempts~\cite{ni2021soar,nikolov2025google,hapim}, most of the existing library migration techniques only focus on finding API mappings\cite{teyton2013automatic, alrubaye2018automating, alrubaye2019use, miningAnalogicalAPIs, zhang2020deep}, without supporting the code transformation.

Large Language Models (LLMs) have shown that they can be effective in various software engineering tasks~\cite{ozkaya2023application, fan2023large, wang2024software}, 
including the necessary upstream tasks for library migration, such as code generation~\cite{nguyen2022empirical, peng2023impact, murali2023codecompose}, code comprehension~\cite{nam2024using}, and code transformation~\cite{wei2023copiloting, xia2023automated, zhou2023hybrid}.
While this suggests that LLMs may be capable of performing library migration, we still need further empirical evidence to support this claim.
Accordingly, this paper investigates the effectiveness of LLMs for library migration for a wide range of libraries and applications.

Our investigation focuses on Python, and we use \migbench \cite{pymigbench}, an existing dataset of \MigAllCount real-world library migrations containing \CCAllCount \codechanges.
The data includes the code before and after the migration, along with detailed descriptions of the \codechanges including labeling of the types of changes using an existing taxonomy \taxonomy~\cite{pymigtax}.

We use three state-of-the-art LLMs: \lm \cite{llamapaper}, \mini \cite{miniref}, and \fouro \cite{fouroref} to perform the migrations.
We refer to these models as \lm, \mn, and \fo in the rest of the paper, respectively.
We evaluate the correctness of the migrations in two ways. First, we compare the LLM-migrated code with the developer-migrated code and report correctness at two granularity levels: the full migration and the individual \ccs. 
We use \taxonomy~\cite{pymigtax} to identify which types of \ccs the LLMs can/cannot handle.
Second, for the subset of migrations with available unit tests, we run the available tests to assess the correctness of the migrated code.

We find that \fo performs the best among the three LLMs, while also being the most expensive.
It correctly migrates at least one \codechange for \FouroMigBMAtLeastOneCorrectPercent of the migrations and fully migrates \FouroMigBMCorrectPercent of the migrations.
At the \cc level, \fo correctly migrates \FouroCCBMTotalCorrectPercent of the \codechanges.
In the test-based evaluation, \FouroMigTestCorrectPercent of the migrations done by \fouro pass the same set of tests that passed in the developer's migration.
\mn, the lower cost alternative of \fo, closely follows \fo in performance
and \lm performs comparatively worst.

Our analysis of the \ccs using \taxonomy reveals that all three LLMs can perform reasonably well on the majority of the \cc types.
Specifically, their performance on higher cardinality  \cc, \ie, changes involving multiple APIs, is promising (\LlamaCCBMHigherCardinalityCorrectPercent, \MiniCCBMHigherCardinalityCorrectPercent and \FouroCCBMHigherCardinalityCorrectPercent correct).
However, all three LLMs struggle with \ccs that require \argTrans, including changes to the argument value or type. 
\fo outperforms the other two models in most of the \cc types, especially in \fcall and higher cardinality \cc{s}.

The evaluation setup above has the advantage of comparing to a developer-performed ground truth migration. However, it uses code that the models may have potentially seen during training, which poses a risk of reciting the migrated code.
To address this, we conduct an additional experiment on \UnseenMigsCount recent well-tested repositories using migrations that never appeared in their commit history.
Overall, we find that even on unseen target code, \lm, \mn, and \fo perfectly migrate \UnseenLlamaAllPassPercent, \UnseenMiniAllPassPercent, and \UnseenFouroAllPassPercent of the migrations respectively, which shows similar relative performance to the previous experiment.

Overall, our study reveals various new opportunities and challenges of using LLMs for library migration.
We find that all three LLMs have high overall correctness, indicating the potential to play a critical role in reducing library migration manual effort.
Also, LLMs are capable of transforming changes that were typically considered difficult, \eg many-to-many code changes and changes between APIs having completely different styles. Challenges include all three LLMs struggling to handle some unit conversions, doing occasional unwanted extra changes, and generalizing exceptions.
Practitioners can use these insights to decide if they want to use LLMs for library migration.
Tool builders can build library migration tools around LLMs by addressing the limitations identified in this study, potentially combining LLMs with traditional program analysis techniques.
The artifact for this study is available at \url{https://figshare.com/articles/conference_contribution/25459000}.

To summarize, our contributions in this paper are as follows:
\begin{enumerate}[leftmargin=*]
    \item We investigate the effectiveness of three LLMs for library migration using an existing dataset of \MigAllCount real-world library migrations, \migbench. 
    \item We evaluate migration correctness in two ways: (1) comparing it to the developer changes at both migration and \cc levels and (2) using unit tests, which provides a run-time evaluation of the migrated code.
    \item Using \taxonomy~\cite{pymigtax}, we identify which types of \ccs the LLMs can/cannot handle.
    \item We verify LLMs' performance on unseen migrations.    
\end{enumerate}

    \section{Background and Terminology}
\label{sec:background}


\subsection{Library Migration}

\textit{Library migration} is the process of updating a software project to replace a used library with another one that provides similar functionality \cite{teyton2012mining}.
The \textit{source} library is the one being replaced, and the \textit{target} library is the one that replaces it \cite{teyton2012mining}.
One \textit{migration} instance refers to a commit in a repository where a migration happened from a specific source library to a target library \cite{pymigbench}, denoted using the notation \pair{source}{target}. 
For example, the commit \href{https://github.com/openstack/ironic/commit/b0607a26}{\texttt{b0607a26}} in repository \href{https://github.com/openstack/ironic}{\textit{openstack/ironic}} 
is a \pair{retrying}{tenacity} migration.

A \textit{\codechange}, or \textit{\cc} for brevity, is a minimal replacement of source library APIs with target library APIs that cannot be meaningfully reduced further without losing the semantics of the change~\cite{pymigbench}.
A migration contains one or more \ccs.
The lines between the boxes in \autoref{fig:mig-example} show \ccs.
For example, segment P1 in \autoref{fig:mig-example-pre} is replaced by segment D1 in \autoref{fig:mig-example-dev}.

We use subscripts \textit{pre}, \textit{dev}, and \textit{llm} to denote data before migration, after developer's migration, and after an LLM's migration, respectively.
For example, \codePre, \codeDev, and \codeLLM denote three states of a code.
\changeDev and \changeLLM denote \ccs by developers and an LLM, respectively.

\subsection{\migbench and \taxonomy}
\migbench is a dataset of real-world Python library migrations \cite{pymigbench}, mined from version control history.
We use the latest available version, version 2.2.5 \cite{pymigtax} in our experiments.
The dataset has \MigAllCount migrations and \CCAllCount \codechanges.
It includes a detailed description of each \cc, including line numbers and API names, which facilitates our evaluation.

\taxonomy is a taxonomy of \codechanges \cite{pymigtax}, built using the first version of \migbench. 
The authors validated its generalizability on additional third-party data.
\taxonomy describes a code change based on three dimensions:
\textit{(1) Program elements}: the types of program elements involved in the change (\eg~\fcall and \attr);
\textit{(2) Cardinality}: how many source APIs are replaced with how many target APIs. For example, \om cardinality means one source API is replaced with multiple target APIs;
    \om, \mo, and \mm cardinalities are commonly referred to as \textit{higher-cardinality}; and
\textit{(3) Properties}: Additional properties to describe the \cc (\eg~\enc when the source and target APIs have different names.). 
The \cc data in \migbench is annotated with the \taxonomy categories.
We use \taxonomy categories (shown in \autoref{tab:rq:cc-freq}) to understand the types of \ccs LLMs can handle.
The \taxonomy paper~\cite{pymigtax} has full category descriptions.

    \section{Experiment Setup}

\subsection{Models}
\label{sec:model-description}
We use the latest available versions of the three LLMs at the time of experiment: 
Meta \llama-70B-Instruct, OpenAI \mini-2024-07-18, and OpenAI \fouro-2024-08-06.
\lm is free and open source with an input+output limit of 8,192 tokens and is trained up until December 2023~\cite{llamadoc,llamapaper}.
The OpenAI models are proprietary, have a output token limit of 16,384 and are trained on data up until October 2023 \cite{openaimodels}.

\subsection{Data Preparation}
Two repositories from \migbench are no longer public, preventing us from using \MigCloneFailedCount migrations.
We clone the remaining repositories containing \MigCloneSuccessCount  migrations.
For each migration commit, we use PyDriller \cite{PyDriller} to extract the content of each file that has recorded \codechanges, both before (\filePre) and after the migration (\fileDev).
We use Python's built-in \texttt{ast.parse} function \cite{docast} to ensure that both \filePre and \fileDev are syntactically valid code and discard \MigSyntaxErrorCount migrations having files with syntax errors.
We conduct our experiments with the remaining \MigExperimentCount migrations containing \CCExperimentCount \codechanges.

\subsection{Migration}
\label{sec:migration}
For each file having \ccs, we use the LLM prompt template in \autoref{fig:prompt} to migrate the code.
The prompt is designed to ensure that the LLM performs focused, controlled migrations while maintaining transparency.
By asking the LLM to explain its changes, we ensure that the LLM justifies its modifications, which can be useful for a human evaluator reviewing the migration. 
The prompt also aims to restrict the LLM from making unrelated modifications such that we remain focused on the task of library migration.

Since LLMs are non-deterministic~\cite{atil2024llmstabilitydetailedanalysis}, we use a temperature of 0 and run each migration 10 times for each model.
We calculate the difference between the runs to understand the variability in the generated code.
We use git-diff~\cite{gitdiff} to compute a diff between each pair of the 10 runs and normalize the number of different lines by dividing it by the total number of lines in \filePre.
We find that on average, two runs on the same file are only \LlamaAvgDiff, \MiniAvgDiff and \FouroAvgDiff different for \lm, \mn, and \fo, respectively.
Given this low variability and the high effort involved in manual validation, we randomly select one run for each model to evaluate the LLMs' results.
Our artifact contains the LLM's migrations for all 10 runs.

    \subsection{Migration Evaluation}

Our goal is to assess the ability of LLMs to migrate Python code between analogous libraries.
However, given that there may be more than one way to correctly migrate an API usage~\cite{chen2020similarapi}, the LLM's migrations might not exactly match the developer changes in \migbench, yet still be correct.
Considering this, we use different strategies to assess correctness through the following research questions.

\def\RQBenchmark{How similar are the LLM migrations to the benchmark migrations?\xspace}
\def\RQTest{How many migrations pass unit tests?\xspace}
\def\RQUnseen{Can LLMs perform migrations they have not seen before?\xspace}

\begin{enumerate}[label={RQ\arabic*},leftmargin=*]
    \item \label{rq:benchmark} \textbf{\RQBenchmark} We consider \changeDev stored in \migbench as the ground truth. We automatically check if the LLM was able to correctly perform all expected changes, while accounting for refactoring and alternative correct changes through manual review.
    \item \label{rq:test} \textbf{\RQTest} 
To evaluate run-time correctness of the migrated code, we use a second evaluation strategy where we run any available unit tests.

    \item \label{rq:unseen} \textbf{\RQUnseen} 
We evaluate on \UnseenReposCount additional well-tested repositories that were updated after the models' training dates, using migrations that never happened in their version-control history. We run the tests to evaluate correctness.
\end{enumerate}

\section{\ref{rq:benchmark} \RQBenchmark}

\subsection{Approach}

\begin{figure*}[t]    
    \includegraphics[width=.98\linewidth]{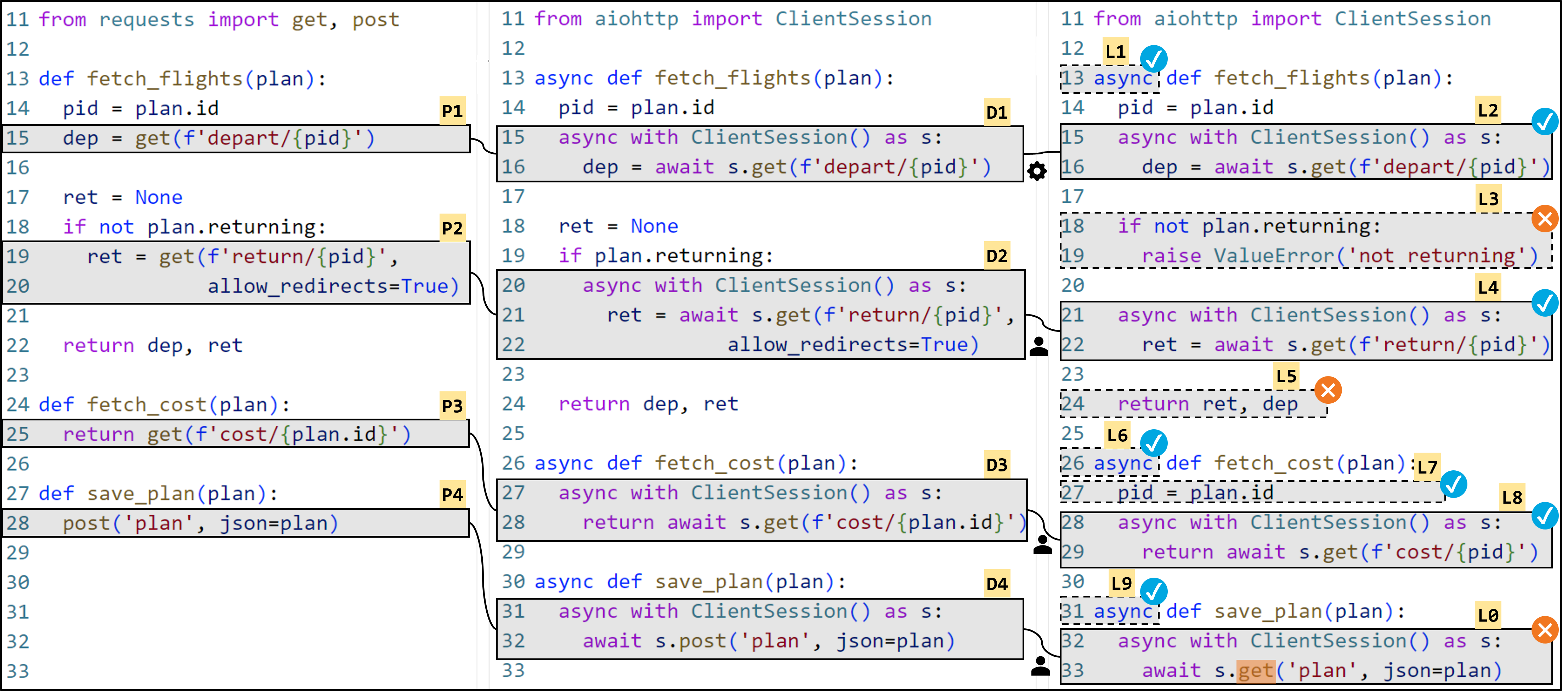}
    
    \begin{subfigure}[t]{.32\textwidth}        
        \vspace{-1.7em}
        \caption{Code before migration (\filePre)}
        \Description{Code before migration}
        \label{fig:mig-example-pre}
    \end{subfigure}
    \begin{subfigure}[t]{.33\textwidth}            
        \vspace{-1.7em}
        \caption{Developer's migration (\fileDev)}
        \Description{Developer's migration}
        \label{fig:mig-example-dev}
    \end{subfigure}
    \begin{subfigure}[t]{.33\textwidth}        
        \vspace{-1.7em}
        \caption{LLM Migration (\fileLLM)}
        \Description{LLM Migration}
        \label{fig:mig-example-llm}
    \end{subfigure}
    
    \caption{
        A sample \pair{requests}{aiohttp} migration.
        The lines between the boxes show matching \codechanges. 
        \faCog~denotes automatic matching, while \faUser~denotes manual matching.
        Dashed lines denote changes that are not recorded in \migbench.
        The blue check marks denote correct changes, while the orange crosses denote incorrect changes.  \vspace{-0.5cm}      
    } 
    \label{fig:mig-example}
\end{figure*}

In this RQ, we assess each LLM's migration correctness by comparing its code changes (\changeLLM) to those made by developers (\changeDev), while manually judging potential alternative changes.
A migration may involve multiple files, each with multiple \codechanges.
We evaluate each \cc individually, and aggregate the results to determine the correctness of the migration.
We now explain the process using the example in \autoref{fig:mig-example} that shows a migration from the library \textit{requests} to \textit{aiohttp}.


\begin{figure}[t!]
    \begin{tcolorbox}[boxrule=1pt, left=2pt, right=2pt, top=2pt, bottom=2pt]
    \footnotesize
        
    The following Python code uses library $<$source-lib$>$.
    Migrate this code to use library $<$target-lib$>$ instead. 
    In the output, first explain the changes you made.
    Then, provide the modified code.
    Do not make any changes to the code that are not
    related to migrating between these two libraries. 
    Do not refactor. Do not reformat. Do not optimize. Do not change coding style. 
    Provided code:  
    \end{tcolorbox}
    \caption{LLM Migration Prompt}\vspace{-0.5cm}
    \label{fig:prompt}
\end{figure}

\subsubsection{Match \ccs}
\label{sec:match-ccs}
The goal of this step is to compare the code changes the LLM made (\changeLLM) with the developer ground truth changes (\changeDev), as well as manually finding alternative correct changes.

\paragraph{Auto change matching}
\label{sec:auto-cc-matching}
We first attempt to automatically match as many changes as possible while ensuring precision \ie not falsely marking a \changeLLM as correct.
Therefore, we consider only \textit{exact} syntactic matches between \changeDev and \changeLLM, ignoring formatting differences.
We find exact matches by identifying the AST nodes related to \changeDev in \fileDev and then matching them to corresponding nodes in \fileLLM.
If there are potentially multiple matches, we check the containing function and use proximity based heuristics demonstrated below.
We ensure that the matched node sets translate to the same code string using \texttt{ast.unparse}~\cite{docast}.

Consider \changeDev \segment{D1} in \autoref{fig:mig-example} which uses \texttt{ClientSession()} and \texttt{get()} from \textit{aiohttp}.
In \fileLLM, these APIs appear in multiple locations, but only Lines 15, 16, 21, and 22 are in the same containing function \texttt{fetch\_flights} as \changeDev.
Based on line number proximity, we identify two node sets \segment{L2} and \segment{L4} as potential matches for this \changeDev.
By comparing the code strings, we find that \segment{L2} exactly matches \changeDev.
By the end of the auto change matching step, we get a set of matched pairs of \changeDev and \changeLLM, a set of unmatched \changeDev (\uChangeDev), and a set of unmatched \changeLLM (\uChangeLLM).

\paragraph{Manual review}
\label{sec:manual-cc-matching}
The manual review step focuses on reviewing \uChangeDev and \uChangeLLM.
We use a three-way diff viewer to manually compare \filePre, \fileDev, and \fileLLM for each file with at least one \uChangeDev or \uChangeLLM.
We use the online API documentation as well as the source code of the source and target libraries to understand the correct API usage in the context of the file.
For \autoref{fig:mig-example}, we manually find that \segment{L4} is a correct alternative to \segment{D2}, noting that the LLM omitted the default value (\texttt{True}) of the \texttt{allow\_redirects} argument, unlike the developer.
\segment{L8} is a correct replacement of \segment{D3}, where the LLM performed a refactoring.
However, we find that the LLM used the function \texttt{get} instead of \texttt{post} at line 33,
making \segment{L0} an incorrect alternative of \segment{D4}.

After trying to match all changes in \changeDev, we may find that some \changeLLM remain unmatched.
Our aim is to categorize these as refactoring (not altering program semantics) or non-refactoring (potentially affecting code behavior).
If a \changeLLM is refactoring, we remove it from \uChangeLLM, as it does not impact migration correctness.
In the example, the LLM introduced handling of a \texttt{ValueError} (\segment{L3}), which changes the program behavior, so we keep it in \uChangeLLM.
Similarly, swapping variables in \segment{L5} is also non-refactoring.

Note that while \migbench records all migration-related \ccs, it does not record additional changes \textit{indirectly} related to the migration, \ie lines that do not have a target library API usage.
For example, adding \texttt{async} to function definitions (\segment{L1}, \segment{L6}, and \segment{L9}) due to async calls introduced by migration are not recorded.
We remove these changes from \uChangeLLM and mark them as correct.
Overall, \segment{L3} and \segment{L5} stay in \uChangeLLM, while \segment{L1}, \segment{L6}, \segment{L7}, and \segment{L9} get removed.

To ensure the quality of the manual review, we have two authors independently review changes.
We first review 194 \ccs from one model, and measure the agreement using Cohen's Kappa \cite{cohen1960coefficient}, but we do not reach substantial agreement.
Therefore, we discuss and resolve disagreements and update our coding guideline.
We then review another 114 \ccs, and achieve a Cohen's Kappa score of \KappaTwo (almost perfect) and \KappaOne (substantial)~ \cite{landis1977measurement}.
Accordingly, we proceed with only one reviewer for the remaining changes.

\subsubsection{Determine migration status}
Based on the matching of \changeDev and \changeLLM, we determine each \cc status as follows (example changes refer to \autoref{fig:mig-example}).

\begin{itemize}[leftmargin=*,topsep=0pt]
    \item \textit{Correct change}: LLM's change is correct, either exactly like the developer (\segment{D1}--\segment{L2}), or with an alternative API (\segment{D2}--\segment{L4}), or with same APIs but with some refactoring (\segment{D3}--\segment{L8}).

    \item \textit{Incorrect change}: The LLM incorrectly implemented the migration change:
    Used an incorrect API (\eg \texttt{get()} instead of \texttt{post()} in \segment{L0}), did not attempt to migrate it at all, or incorrectly removed part of a code.
\end{itemize}

Based on the individual matched code changes, we now automatically determine an overall migration's status:
\begin{itemize}[leftmargin=*,topsep=0pt,parsep=0pt]
    \item \textit{\MigStatusResponseFailure}: The LLM did not generate a \fileLLM for at least one \filePre
(e.g., due to token limit or API timeout).        

    \item \textit{\MigStatusSyntaxError}: The LLM generated a \fileLLM for all \filePre, but at least one \fileLLM has syntax errors. 

    \item \textit{\MigStatusIncorrect}: The LLM could not correctly migrate \textit{any} of the changes marked in \migbench.
    We assign this status when all \changeDev across the migration remain unmatched.

    \item \textit{\MigStatusPartialCorrect}: The LLM correctly migrated only \textit{some} of \changeDev.
    We assign this status when there are only some unmatched \changeDev.

    \item \textit{\MigStatusExtraChange}: The LLM correctly migrated \textit{all} \changeDev but also performed some non-refactoring changes. We assign this status when there are no unmatched \changeDev in any file, but there are some remaining unmatched \changeLLM.

    \item \textit{\MigStatusCorrect}: The LLM correctly migrated \textit{all} \changeDev for this migration, without any non-refactoring changes.    
    We assign this status when there are no unmatched \changeDev or \changeLLM left in any of the files after the matching process.
\end{itemize}
    \subsection{Findings: Migration level correctness}

\autoref{tab:bm-eval-mig-level} shows the migration level results.
We find that \fo performs the best with \FouroMigBMAtLeastOneCorrectPercent of the migrations having at least one correct \cc, closely followed by \mn with \MiniMigBMAtLeastOneCorrectPercent.
When considering fully correct migrations, \fo outperforms \mn considerably, with \FouroMigBMCorrectPercent compared to \MiniMigBMCorrectPercent.
\lm has the lowest performance, with only \LlamaMigBMCorrectPercent being fully correct and \LlamaMigBMAtLeastOneCorrectPercent having at least one correct \cc.
This is mainly due to a higher proportion of \textit{\MigStatusResponseFailureLower}s where \lm ran into token limits.
\lm also produces more syntax errors compared to the other two models (\LlamaMigBMSyntaxErrorPercent vs. \MiniMigBMSyntaxErrorPercent and \FouroMigBMSyntaxErrorPercent).

\begin{findingenv}{\rqBench, Migration Level}{finding:mig-correctness}
  All three LLMs were able to perform correct migrations with \fo performing best: \FouroMigBMAtLeastOneCorrectPercent of its migrations were partially correct and \FouroMigBMCorrectPercent were fully correct.
\end{findingenv}

\subsection{Findings: \cc level correctness}

\begin{table}[t]
    \renewcommand{\arraystretch}{1}
    \centering
    \caption{Correctness of \migbench migrations (RQ1, migration level)}
    \scriptsize
    \label{tab:bm-eval-mig-level}
    {%
    \begin{tabular}{>{\quad}lrrr}
    \toprule
    ~                         & \multicolumn{3}{c}{Number (percentage) of migrations}\\ 
    \cmidrule{2-4}
    \hspace{-.8em}Status       & {\llama} & {\mini} & {\fouro} \\
    \midrule  
    \hspace{-.8em}\textit{At least partially correct} \\
    \MigStatusCorrect          & \LlamaMigBMCorrectCount~(\LlamaMigBMCorrectPercent) & \MiniMigBMCorrectCount~(\MiniMigBMCorrectPercent) & \FouroMigBMCorrectCount~(\FouroMigBMCorrectPercent) \\
    Correct w/ non-refactorings      & \LlamaMigBMCorrectWithExtraChangesCount~(\LlamaMigBMCorrectWithExtraChangesPercent) & \MiniMigBMCorrectWithExtraChangesCount~(\MiniMigBMCorrectWithExtraChangesPercent) & \FouroMigBMCorrectWithExtraChangesCount~(\FouroMigBMCorrectWithExtraChangesPercent) \\
    \MigStatusPartialCorrect   & \LlamaMigBMPartiallyCorrectCount~(\LlamaMigBMPartiallyCorrectPercent) & \MiniMigBMPartiallyCorrectCount~(\MiniMigBMPartiallyCorrectPercent) & \FouroMigBMPartiallyCorrectCount~(\FouroMigBMPartiallyCorrectPercent) \\    
    \textit{Subtotal}          & \textit{\LlamaMigBMAtLeastOneCorrectCount~(\LlamaMigBMAtLeastOneCorrectPercent)} & \textit{\MiniMigBMAtLeastOneCorrectCount~(\MiniMigBMAtLeastOneCorrectPercent)} & \textit{\FouroMigBMAtLeastOneCorrectCount~(\FouroMigBMAtLeastOneCorrectPercent)} \\
    \midrule
    \hspace{-.8em}\textit{Fully incorrect} \\
    \MigStatusIncorrect        & \LlamaMigBMIncorrectCount~(\LlamaMigBMIncorrectPercent) & \MiniMigBMIncorrectCount~(\MiniMigBMIncorrectPercent) & \FouroMigBMIncorrectCount~(\FouroMigBMIncorrectPercent) \\
    \MigStatusSyntaxError      & \LlamaMigBMSyntaxErrorCount~(\LlamaMigBMSyntaxErrorPercent) & \MiniMigBMSyntaxErrorCount~(\MiniMigBMSyntaxErrorPercent) & \FouroMigBMSyntaxErrorCount~(\FouroMigBMSyntaxErrorPercent) \\
    \MigStatusResponseFailure  & \LlamaMigBMResponseFailureCount~(\LlamaMigBMResponseFailurePercent) & \MiniMigBMResponseFailureCount~(\MiniMigBMResponseFailurePercent) & \FouroMigBMResponseFailureCount~(\FouroMigBMResponseFailurePercent) \\    
    \textit{Subtotal}          & \textit{\LlamaMigBMNoCorrectCount~(\LlamaMigBMNoCorrectPercent)} & \textit{\MiniMigBMNoCorrectCount~(\MiniMigBMNoCorrectPercent)} & \textit{\FouroMigBMNoCorrectCount~(\FouroMigBMNoCorrectPercent)} \\
    \midrule
    Total migrations    & \MigExperimentCount (100\%) & \MigExperimentCount (100\%) & \MigExperimentCount (100\%) \\ 
    \bottomrule
    \end{tabular}
    }\vspace{-0.5cm}
\end{table}

We now present the correctness results at the individual \cc level.
The \MigExperimentCount migrations we use for our evaluation contain a total of \CCExperimentCount \ccs.
However, when there is a response failure or syntax error in the LLM's migration, we cannot evaluate the correctness of the corresponding \ccs in that file.
Accordingly, in this analysis level, we exclude \LlamaCCBMExcludedCount \ccs from \lm, \MiniCCBMExcludedCount from \mn, and \FouroCCBMExcludedCount from \fo.
This leaves us with \LlamaCCBMEvaluatedCount, \MiniCCBMEvaluatedCount, and \FouroCCBMEvaluatedCount \ccs to evaluate for the three LLMs, respectively.

\subsubsection{Overall code change correctness}

\autoref{tab:cc-level-summary} shows the overall correctness of the LLMs' migrations at the \cc level.
\fo performs best, with \FouroCCBMTotalCorrectPercent of the \ccs being correct.
Interestingly, \mn and \lm both perform equally well at the \cc level, with both having \MiniCCBMTotalCorrectPercent correct \ccs.
This is in contrast to the migration level, where \mn performs better than \lm (\MiniMigBMCorrectPercent vs. \LlamaMigBMCorrectPercent).
This suggests that while \mn attempted more \ccs compared to \lm, the ones that \lm was able to attempt without failures or syntax errors were comparatively correct.

\subsubsection{Code change correctness by category}

To understand if some types of \ccs are more difficult to migrate, \autoref{tab:rq:cc-freq} shows the distribution of \cc correctness across the different \taxonomy categories.
The second column shows how often each \taxonomy category appears in the \CCAllCount \ccs in \migbench, which provides perspective on the category's frequency in practice. 
For example, \FCallTotalCount(\FCallTotalPercent) of all \CCAllCount \ccs involve \fcall{s}.
The next three column groups show the performance of each model.
Since each LLM has a different number of evaluated code changes,
we show the number of evaluated code changes from each category (\evalCountCell) alongside the proportion of these code changes that are correct (\correctPercentCell).
For example, the \textit{\Fcall} row under the \textit{Program elements} group shows that \lm's syntactically correct migrations included \LlamaCCBMFCallTotalCount \ccs that involve \fcall{s}, and that \LlamaCCBMFCallCorrectPercent of these \ccs were correctly migrated.

\begin{table}[t!]
    \renewcommand{\arraystretch}{1}
    \centering
    \caption{Correctness of \migbench \codechanges (RQ1, overall code change level)}
    \scriptsize
    \label{tab:cc-level-summary}
    {%
    \begin{tabular}{lrrr}
    \toprule
    ~                         & \multicolumn{3}{c}{Number (percentage) of \ccs}\\ 
    \cmidrule{2-4}
    Status       & {\llama} & {\mini} & {\fouro} \\
    \midrule    
    \CCStatusCorrect            & \LlamaCCBMTotalCorrectCount~(\LlamaCCBMTotalCorrectPercent) & \MiniCCBMTotalCorrectCount~(\MiniCCBMTotalCorrectPercent) & \FouroCCBMTotalCorrectCount~(\FouroCCBMTotalCorrectPercent) \\
    \CCStatusIncorrect            & \LlamaCCBMTotalIncorrectCount~(\LlamaCCBMTotalIncorrectPercent) & \MiniCCBMTotalIncorrectCount~(\MiniCCBMTotalIncorrectPercent) & \FouroCCBMTotalIncorrectCount~(\FouroCCBMTotalIncorrectPercent) \\
    \midrule
    Evaluated \ccs & \LlamaCCBMEvaluatedCount~(100\%) & \MiniCCBMEvaluatedCount~(100\%) & \FouroCCBMEvaluatedCount~(100\%) \\
    Excluded \ccs  & \LlamaCCBMExcludedCount & \MiniCCBMExcludedCount & \FouroCCBMExcludedCount \\
    \midrule
    \hspace{-.8em}Total \ccs     & \CCExperimentCount & \CCExperimentCount & \CCExperimentCount \\
    \bottomrule
    \end{tabular}\vspace{-0.2cm}
    }
\end{table}

\paragraph{Program elements}
\Fcall{s} are the most common program elements in the \migbench (\FCallTotalPercent frequency).
All three LLMs correctly migrate a high proportion of these changes, with  \LlamaCCBMFCallCorrectPercent, \MiniCCBMFCallCorrectPercent, and \FouroCCBMFCallCorrectPercent of the \fcall{s} being correctly migrated  by the three LLMs. Almost all \textit{migrations} in \migbench require migrating \imp statements, a task that all three LLMs perform well, especially \fo correctly migrates nearly all of them (\FouroCCBMImpCorrectPercent).

\begin{table}[t]
    
    \newcommand{\cellSpace}{\hspace{.5em}}
    \newcommand{\negCellSpace}{\hspace{-.5em}}

    \centering
    \caption{Correctness of \ccs across \taxonomy categories (\rqBench, code change level by category).    
    }
    \label{tab:rq:cc-freq}
    \scriptsize
    {
    \begin{tabular}{@{}>{\cellSpace}l@{~}rr@{~}rr@{~}rr@{~}r@{}}
    \toprule
    \multirow[b]{2}{*}{\makecell{\negCellSpace\taxonomy\\ category}}             & \multirow[b]{2}{*}{\makecell{Frequency in \\ \migbench}} & \multicolumn{2}{c}{\llama}                                                                                    & \multicolumn{2}{c}{\mini}                                                                                  & \multicolumn{2}{c}{\fouro}                                                                                      \\
    \cmidrule(l){3-4}\cmidrule(l){5-6}\cmidrule(l){7-8}
     &  
    & \evalCountCell          
    & \correctPercentCell
    & \evalCountCell 
    & \correctPercentCell                                         
    & \evalCountCell 
    & \correctPercentCell                                        
    \\
    \midrule
    \negCellSpace{\textit{Program elements}}        \\
    \Fcall
    & \FCallTotalCount (\FCallTotalPercent)
    & \LlamaCCBMFCallTotalCount
    & \LlamaCCBMFCallCorrectPercent 
    & \MiniCCBMFCallTotalCount 
    & \MiniCCBMFCallCorrectPercent  
    & \FouroCCBMFCallTotalCount
    & \FouroCCBMFCallCorrectPercent 
    \\
    \Imp       
    & \ImpTotalCount (\ImpTotalPercent)
    & \LlamaCCBMImpTotalCount   
    & \LlamaCCBMImpCorrectPercent  
    & \MiniCCBMImpTotalCount
    & \MiniCCBMImpCorrectPercent 
    & \FouroCCBMImpTotalCount
    & \FouroCCBMImpCorrectPercent 
    \\
    \Dec       
    & \DecTotalCount (\DecTotalPercent)
    & \LlamaCCBMDecTotalCount
    & \LlamaCCBMDecCorrectPercent 
    & \MiniCCBMDecTotalCount
    & \MiniCCBMDecCorrectPercent 
    & \FouroCCBMDecTotalCount
    & \FouroCCBMDecCorrectPercent 
    \\
    \Attr      
    & \AttrTotalCount (\AttrTotalPercent)
    & \LlamaCCBMAttrTotalCount
    & \LlamaCCBMAttrCorrectPercent 
    & \MiniCCBMAttrTotalCount
    & \MiniCCBMAttrCorrectPercent 
    & \FouroCCBMAttrTotalCount
    & \FouroCCBMAttrCorrectPercent 
    \\
    \Type      
    & \TypeTotalCount (\TypeTotalPercent)
    & \LlamaCCBMTypeTotalCount
    & \LlamaCCBMTypeCorrectPercent 
    & \MiniCCBMTypeTotalCount
    & \MiniCCBMTypeCorrectPercent 
    & \FouroCCBMTypeTotalCount
    & \FouroCCBMTypeCorrectPercent 
    \\
    \Exc       
    & \ExcTotalCount (\ExcTotalPercent)
    & \LlamaCCBMExcTotalCount
    & \LlamaCCBMExcCorrectPercent 
    & \MiniCCBMExcTotalCount
    & \MiniCCBMExcCorrectPercent 
    & \FouroCCBMExcTotalCount
    & \FouroCCBMExcCorrectPercent 
    \\
    \fRefCell      
    & \FRefTotalCount (\FRefTotalPercent)
    & \LlamaCCBMFRefTotalCount
    & \LlamaCCBMFRefCorrectPercent 
    & \MiniCCBMFRefTotalCount
    & \MiniCCBMFRefCorrectPercent 
    & \FouroCCBMFRefTotalCount
    & \FouroCCBMFRefCorrectPercent 
    \\
    
    \midrule
    \negCellSpace{\textit{Properties}}        \\
    \NoProps      
    & \NoPropsTotalCount (\NoPropsTotalPercent)
    & \LlamaCCBMNoPropsTotalCount
    & \LlamaCCBMNoPropsCorrectPercent 
    & \MiniCCBMNoPropsTotalCount
    & \MiniCCBMNoPropsCorrectPercent 
    & \FouroCCBMNoPropsTotalCount
    & \FouroCCBMNoPropsCorrectPercent 
    \\
    \elemNCCell       
    & \ElemNCTotalCount (\ElemNCTotalPercent)
    & \LlamaCCBMElemNCTotalCount 
    & \LlamaCCBMElemNCCorrectPercent 
    & \MiniCCBMElemNCTotalCount 
    & \MiniCCBMElemNCCorrectPercent 
    & \FouroCCBMElemNCTotalCount 
    & \FouroCCBMElemNCCorrectPercent 
    \\
    \argAddCell       
    & \ArgAddTotalCount (\ArgAddTotalPercent)
    & \LlamaCCBMArgAddTotalCount 
    & \LlamaCCBMArgAddCorrectPercent
    & \MiniCCBMArgAddTotalCount 
    & \MiniCCBMArgAddCorrectPercent 
    & \FouroCCBMArgAddTotalCount 
    & \FouroCCBMArgAddCorrectPercent 
    \\
    \argDelCell   
    & \ArgDelTotalCount (\ArgDelTotalPercent)
    & \LlamaCCBMArgDelTotalCount 
    & \LlamaCCBMArgDelCorrectPercent 
    & \MiniCCBMArgDelTotalCount 
    & \MiniCCBMArgDelCorrectPercent 
    & \FouroCCBMArgDelTotalCount 
    & \FouroCCBMArgDelCorrectPercent 
    \\
    \argTransCell     
    & \ArgTransTotalCount (\ArgTransTotalPercent)
    & \LlamaCCBMArgTransTotalCount 
    & \LlamaCCBMArgTransCorrectPercent 
    & \MiniCCBMArgTransTotalCount 
    & \MiniCCBMArgTransCorrectPercent 
    & \FouroCCBMArgTransTotalCount 
    & \FouroCCBMArgTransCorrectPercent 
    \\
    \paramAddCell 
    & \ParamAddTotalCount (\ParamAddTotalPercent)
    & \LlamaCCBMParamAddTotalCount 
    & \LlamaCCBMParamAddCorrectPercent 
    & \MiniCCBMParamAddTotalCount 
    & \MiniCCBMParamAddCorrectPercent 
    & \FouroCCBMParamAddTotalCount 
    & \FouroCCBMParamAddCorrectPercent 
    \\
    \argNCCell        
    & \ArgNCTotalCount (\ArgNCTotalPercent)
    & \LlamaCCBMArgNCTotalCount 
    & \LlamaCCBMArgNCCorrectPercent 
    & \MiniCCBMArgNCTotalCount 
    & \MiniCCBMArgNCCorrectPercent 
    & \FouroCCBMArgNCTotalCount 
    & \FouroCCBMArgNCCorrectPercent 
    \\
    \asyncTransCell   
    & \AsyncTransTotalCount (\AsyncTransTotalPercent)
    & \LlamaCCBMAsyncTransTotalCount 
    & \LlamaCCBMAsyncTransCorrectPercent 
    & \MiniCCBMAsyncTransTotalCount 
    & \MiniCCBMAsyncTransCorrectPercent 
    & \FouroCCBMAsyncTransTotalCount 
    & \FouroCCBMAsyncTransCorrectPercent 
    \\
    \outTransCell     
    & \OutTransTotalCount (\OutTransTotalPercent)
    & \LlamaCCBMOutTransTotalCount 
    & \LlamaCCBMOutTransCorrectPercent 
    & \MiniCCBMOutTransTotalCount 
    & \MiniCCBMOutTransCorrectPercent 
    & \FouroCCBMOutTransTotalCount 
    & \FouroCCBMOutTransCorrectPercent 
    \\

    \midrule
    \negCellSpace{\textit{Cardinality}}        \\
    \OO        
    & \OOTotalCount (\OOTotalPercent)
    & \LlamaCCBMOOTotalCount 
    & \LlamaCCBMOOCorrectPercent 
    & \MiniCCBMOOTotalCount 
    & \MiniCCBMOOCorrectPercent 
    & \FouroCCBMOOTotalCount 
    & \FouroCCBMOOCorrectPercent 
    \\
    \OZ        
    & \OZTotalCount (\OZTotalPercent)
    & \LlamaCCBMOZTotalCount 
    & \LlamaCCBMOZCorrectPercent 
    & \MiniCCBMOZTotalCount 
    & \MiniCCBMOZCorrectPercent 
    & \FouroCCBMOZTotalCount 
    & \FouroCCBMOZCorrectPercent 
    \\
    \ZO        
    & \ZOTotalCount (\ZOTotalPercent)
    & \LlamaCCBMZOTotalCount 
    & \LlamaCCBMZOCorrectPercent 
    & \MiniCCBMZOTotalCount 
    & \MiniCCBMZOCorrectPercent 
    & \FouroCCBMZOTotalCount 
    & \FouroCCBMZOCorrectPercent 
    \\
    \textit{Higher Cardinality} 
    & 
    \textit{\HigherCardinalityTotalCount (\HigherCardinalityTotalPercent)}
    & \textit{\LlamaCCBMHigherCardinalityTotalCount} 
    & \textit{\LlamaCCBMHigherCardinalityCorrectPercent} 
    & \textit{\MiniCCBMHigherCardinalityTotalCount} 
    & \textit{\MiniCCBMHigherCardinalityCorrectPercent} 
    & \textit{\FouroCCBMHigherCardinalityTotalCount} 
    & \textit{\FouroCCBMHigherCardinalityCorrectPercent} 
    \\
    \OM
    & \OMTotalCount (\OMTotalPercent)
    & \LlamaCCBMOMTotalCount 
    & \LlamaCCBMOMCorrectPercent 
    & \MiniCCBMOMTotalCount 
    & \MiniCCBMOMCorrectPercent 
    & \FouroCCBMOMTotalCount     
    & \FouroCCBMOMCorrectPercent 
    \\
    \MO        
    & \MOTotalCount (\MOTotalPercent)
    & \LlamaCCBMMOTotalCount 
    & \LlamaCCBMMOCorrectPercent 
    & \MiniCCBMMOTotalCount 
    & \MiniCCBMMOCorrectPercent 
    & \FouroCCBMMOTotalCount 
    & \FouroCCBMMOCorrectPercent 
    \\
    \MM        
    & \MMTotalCount (\MMTotalPercent)
    & \LlamaCCBMMMTotalCount 
    & \LlamaCCBMMMCorrectPercent 
    & \MiniCCBMMMTotalCount 
    & \MiniCCBMMMCorrectPercent 
    & \FouroCCBMMMTotalCount 
    & \FouroCCBMMMCorrectPercent 
    \\
    \midrule
    \negCellSpace{}Total 
    & \CCAllCount 
    & \LlamaCCBMEvaluatedCount 
    & \LlamaCCBMTotalCorrectPercent
    & \MiniCCBMEvaluatedCount  
    & \MiniCCBMTotalCorrectPercent 
    & \FouroCCBMEvaluatedCount  
    & \FouroCCBMTotalCorrectPercent 
    \\
    \bottomrule
    \end{tabular}
    }\vspace{-0.6cm}
\end{table}


\mn slightly outperforms the other models in migrating \dec{s} (\MiniCCBMDecCorrectPercent vs. \LlamaCCBMDecCorrectPercent and \FouroCCBMDecCorrectPercent); however, it does worst in \attr{s} (\MiniCCBMAttrCorrectPercent vs. \LlamaCCBMAttrCorrectPercent and \FouroCCBMAttrCorrectPercent).
\fo performs better than other models in migrating \type{s}, \exc{s} and \fref{s}, followed by \mn.
\lm and \mn both struggle with migrating \fref{s} (\LlamaCCBMFRefCorrectPercent and \MiniCCBMFRefCorrectPercent correct), while \fo correctly performs all such \ccs.
While this last category shows most discrepancy between the models, this program element is present in only \FRefTotalPercent of the \ccs, therefore does not affect the overall performance significantly.

\paragraph{Properties}
The property \textit{\noProps} indicates that the source and target APIs are identical, and hence no changes are required.
While \fo correctly leaves almost all of these APIs unchanged (\FouroCCBMNoPropsCorrectPercent),
\lm and \mn incorrectly changed several of them, resulting in \LlamaCCBMNoPropsCorrectPercent and \MiniCCBMNoPropsCorrectPercent correct, respectively.

\textit{\ElemNC} is the most common property in \migbench.
This indicates that the target APIs commonly bear different names from the source APIs. 
All LLMs perform well in making this change, with \lm, \mn, and \fo correctly migrating \LlamaCCBMElemNCCorrectPercent, \MiniCCBMElemNCCorrectPercent, and \FouroCCBMElemNCCorrectPercent changes, respectively.

All three LLMs performed better or equally well in \textit{\argDel} compared to those with \textit{\argAdd}, which is expected as adding an argument requires the LLM to find a suitable replacement, while deleting an argument is a simpler task.
The \textit{\ArgTrans} property indicates that an argument requires various changes, such as changing the type or value.
The three LLMs correctly migrate only 76-77\% of the \argTrans changes.
We discuss specific failure instances in \autoref{sec:discussion}.

The property \textit{\paramAdd} is found in migrations from or to the click library \cite{libclick}, where some decorators require adding a parameter to the decorated function.
All three LLMs perform well in migrating these changes; interestingly, \mn performs better than \fo (\MiniCCBMParamAddCorrectPercent vs. \FouroCCBMParamAddCorrectPercent).
\mn also performed better than \fo in migrating \dec{s}, which are common in the click library.
These are the only two categories where \fo does not perform the best.

\textit{\ArgNC} applies when an argument representing the same semantics has different names in the source and target APIs.
While \fo performs reasonably well in migrating these changes (\FouroCCBMArgNCCorrectPercent), \lm and \mn fail frequently with them (only \LlamaCCBMArgNCCorrectPercent and \MiniCCBMArgNCCorrectPercent correct).

\paragraph{Cardinality} 
\OO \ccs are the most common in \migbench, and all three LLMs perform well in migrating them.
\OZ \ccs are the ones where a source API needs to be removed, but no target needs to be added.
Because it is a simple delete operation, all three LLMs migrate all or almost all of these changes correctly.
Its counterpart, \ZO \ccs, are the ones where a target needs to be added, but no source needs to be removed.
Interestingly, \lm performs the best in migrating these changes (\LlamaCCBMZOCorrectPercent correct), making this the only category where it performs better than the other models (\MiniCCBMZOCorrectPercent and \FouroCCBMZOCorrectPercent correct).
While previous research showed that higher-cardinality \ccs are generally difficult compared to \oo \ccs \cite{alrubaye2018automating,wang2016transforming, zhang2020deep,HuangMappingAPI2024},
the \migbench dataset shows that higher cardinality changes are frequent with \HigherCardinalityLibPairPercent of the library pairs requiring at least one higher-cardinality \cc \cite{pymigtax}.
Overall, we find that the LLMs are reasonably successful at migrating them, with \LlamaCCBMHigherCardinalityCorrectPercent, \MiniCCBMHigherCardinalityCorrectPercent, and \FouroCCBMHigherCardinalityCorrectPercent correctness rates.

\begin{findingenv}{\rqBench, Code Change Level}{finding:cc-correctness}
    LLMs correctly migrate a high proportion of most \ccs, with \fo achieving \FouroCCBMTotalCorrectPercent correct \ccs.
    The models perform reasonably well in migrating difficult higher-cardinality \ccs (\fo gets \FouroCCBMHigherCardinalityCorrectPercent correct), but struggle relatively more with \argTrans and \argNC.
    
\end{findingenv}

    \section{\ref{rq:test} \RQTest}

\subsection{Approach}
\label{sec:unittest-approach}
We now assess the same LLM \migbench migrations from RQ1 but using the unit tests available in the corresponding repositories.
A test-based evaluation that runs the code ensures that the migration achieves the expected behavior.
Note that we should not expect that an LLM migration improves the rate of passing tests;  a successful migration means that each test that previously passed on \codeDev must still pass on \codeLLM.
We now explain our evaluation set up.

\subsubsection{Preparing the code}
We make a copy of the repository after the developers' migration (\codeDev).
We then make another copy of developer's migration, but replace all \fileDev with \fileLLM; we refer to this as \codeLLM.
The idea is that \codeLLM is basically the version of \codeDev where the LLM did the migration on behalf of the developer.

\subsubsection{Setting up the virtual environment}
To set up a virtual environment, we need to identify the Python version used in the client repository.
If this information is available in the \texttt{setup.py} file, we use that version.
Otherwise, we resolve the version based on the migration commit date as follows.
We first identify the release dates of all minor versions of Python (3.6, 3.7 etc).
Then, we find the latest release date that is before the migration commit date.
This is the latest Python version that was available at the time of the migration; we use this version to create the virtual environment.
Next, we install the code dependencies using \texttt{pyproject.toml}, \texttt{setup.py} and requirements files.
In cases where specific dependency versions are not specified in those files, we look up the version history of the dependency on PyPI and install the latest version available at the migration commit date, similar to how we resolve the Python version. This ensures that the dependencies are compatible with the code at the time of migration.

\subsubsection{Running the tests and coverage}
Once the virtual environment is ready, we run all the unit tests while measuring coverage in the repository on \codeDev. 
If the run has errors, we read the error log and try to fix the errors, and run the tests again.
We maintain a configuration file where we record any project-specific configurations or commands we used. 
We include this information in our artifact.

We find that \MigHasTestsCount out of the \MigExperimentCount migrations have at least one test.
Among these, \MigErrorRunningTestCount migrations had unresolvable errors. 
The failures are primarily due to missing dependencies, specially for older projects.
The remaining \MigTestExecutedCount migrations have tests that we are able to successfully run on \codeDev.

However, for the tests to be useful to validate migration, they must cover the \codechanges.
Using the coverage report, we find that only \MigCoveredCount migrations have at least one test that covers the code changes, and among them, \MigCoveredPassingCount migrations have at least one test that passes on \codeDev.
We run the entire test suite on \codeLLM for these \MigCoveredPassingCount migrations for each model using the same randomly selected LLM run used in \rqBench.

\subsubsection{Determining migration status}
We compare the test results between \codeDev and \codeLLM and assign the following statuses to the migrations:
\begin{itemize}[leftmargin=*]
    \item \textit{\MigStatusCorrect}: All passing tests in \codeDev also pass in \codeLLM.
    \item \textit{\MigStatusPartialCorrect}: Some of the tests that pass in \codeDev pass in \codeLLM, but the others fail or raise run-time errors.
    \item \textit{\MigStatusIncorrect}: None of the passing tests in \codeDev pass in \codeLLM; they all fail or raise run-time errors.   
\end{itemize}

\subsection{Findings}
\label{sec:utest-eval-findings}

\begin{table}[t]
    \centering
    \caption{Correctness of \MigCoveredPassingCount \migbench migrations using their available unit tests (RQ2)}
    \scriptsize
    \label{tab:utest-result}

    \begin{tabular}{lrrr}
    \toprule
    ~                         & \multicolumn{3}{c}{Number (percentage) of correct migrations}\\ 
    
    \cmidrule{2-4}
    Status                    & \llama                                                    & \mini & \fouro                                                                                \\
    \midrule  
    \MigStatusCorrect         & \LlamaMigTestCorrectCount                (\LlamaMigTestCorrectPercent)                & \MiniMigTestCorrectCount                 (\MiniMigTestCorrectPercent)                   & \FouroMigTestCorrectCount                 (\FouroMigTestCorrectPercent)                   \\
    \MigStatusPartialCorrect  & \LlamaMigTestPartiallyCorrectCount       (\LlamaMigTestPartiallyCorrectPercent)       & \MiniMigTestPartiallyCorrectCount        (\MiniMigTestPartiallyCorrectPercent)          & \FouroMigTestPartiallyCorrectCount        (\FouroMigTestPartiallyCorrectPercent)          \\
    \MigStatusIncorrect       & \LlamaMigTestIncorrectCount              (\LlamaMigTestIncorrectPercent)              & \MiniMigTestIncorrectCount               (\MiniMigTestIncorrectPercent)                 & \FouroMigTestIncorrectCount               (\FouroMigTestIncorrectPercent)                 \\    
    \midrule
    Evaluated       & \MigTestEvaluatedCount               & \MigTestEvaluatedCount & \MigTestEvaluatedCount \\    
    \bottomrule
    \end{tabular}\vspace{-0.2cm}
\end{table}

\autoref{tab:utest-result} shows the results for the \MigTestEvaluatedCount migrations. 
\fo has the highest percentage of correct migrations, with \FouroMigTestCorrectPercent, followed by \mn with \MiniMigTestCorrectPercent, and \lm with \LlamaMigTestCorrectPercent.
\mn has more partially correct migrations than \fo and \lm, leading it to have an equal number of incorrect migrations as \fo (\MiniMigTestIncorrectPercent).
\lm, on the other hand, has a much higher proportion of incorrect migrations (\LlamaMigTestIncorrectPercent).

\begin{findingenv}{\rqTest}{finding:test-correctness}
Out of \MigTestEvaluatedCount \migbench migrations with unit tests covering the \codechanges, the LLMs correctly migrated \LlamaMigTestCorrectPercent-\FouroMigTestCorrectPercent, with \fo being the highest.  
\end{findingenv}

\section{\ref{rq:unseen} \RQUnseen}
\subsection{Approach}
All the migrations in \migbench happened prior to the known training cutoff date of the three models (Late 2023).
Therefore, there is a possibility that the LLMs are simply reciting the migrated code they have seen before (for this particular code base).
To validate whether LLMs are able to perform migrations that are not known to them, but for libraries they are aware of, 
we run an experiment with repositories that never contained the target migration and that are updated after the training cutoff dates. 
To allow validating the migration, we choose repositories with high test coverage.

We use SEART~\cite{seart} to find Python repositories that are updated on or after January 1, 2024.
We only keep repositories that have a \texttt{requirements.txt} file to allow us to set up the environment to run tests.
We clone the latest version of the repositories one by one, set up a virtual environment using the requirements file, and then run tests with coverage.
We stop after we find \UnseenReposCount repositories whose tests pass and achieve at least 95\% statement coverage. 

For each repository, we select a library listed in the requirements file and identify an analogous library that provides similar functionality through an online search. 
Overall, we choose \UnseenLibPairsCount unique library pairs, \UnseenOldLibPairsCount from \migbench and \UnseenNewLibPairsCount external to it.
This allows us to evaluate migration of unseen code, both on library pairs we have seen the models perform on before, as well as on new pairs.
To ensure that the models have never seen the expected code that uses the target library in these repositories, we traverse the commit histories of each repository and confirm that the target library never appeared in the requirements file. 
This also removes the possibility that the LLMs have previously seen an inverse migration from the target library to the source library in the repository's history.

Next, we find the code files that use the source library and thus require migration.
For each of the \UnseenReposCount repositories, we parse each file using the Python AST module to find the files that import the source library and thus require migration.
We also manually locate the API usages in those files to confirm that it does not just import the library but also uses it.
We then run the available tests to record their current status and also manually verify that the API usages of the source libraries are covered by the tests.
Finally, we migrate each of the identified files using the same prompt template we used before (\autoref{fig:prompt}).
We run the tests after the migration and compare the test results to those before the migration.

Overall, for this experiment, we have a total of \UnseenMigsCount migrations in \UnseenReposCount repositories between \UnseenLibPairsCount unique library pairs.
The repositories have between \UnseenMinFilesCount and \UnseenMaxFilesCount migration-related files (median \UnseenMedianFilesCount) and between \UnseenMinApisCount to \UnseenMaxApisCount source API usages that require migration (median \UnseenMedianApisCount).
The number of tests in the 10 repositories ranges from \UnseenMinTestsCount to \UnseenMaxTestsCount (median \UnseenMedianTestsCount), all passing before migration with 95\% or more statement coverage.

    \subsection{Findings}
\label{sec:unseen-eval-findings}

\begin{figure}[t]
    \centering
    \includegraphics[width=1\columnwidth]{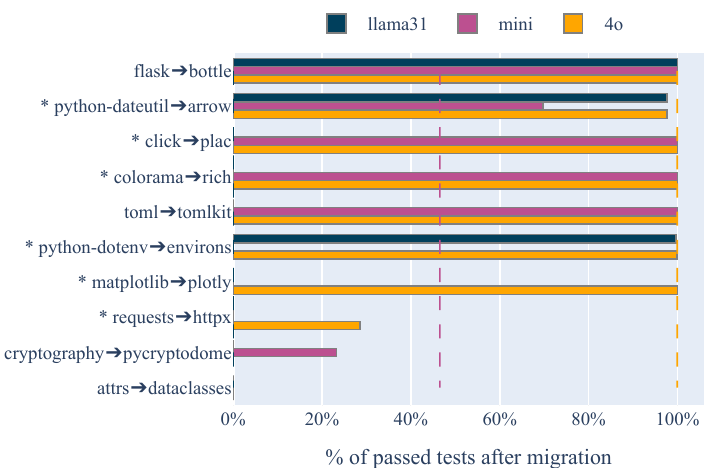} \\    
    \caption{Correctness of migrations unseen by the LLMs (RQ3). Asterisks (\textasteriskcentered) indicate the library pair is not in \migbench.}
    \label{fig:unseen-result}
    \vspace{-5mm}
\end{figure}

The migrations have between \UnseenMinModifiedLines and \UnseenMaxModifiedLines lines modified per migration (median \UnseenMedianModifiedLines), confirming that the LLMs attempted to migrate the code rather than returning the same code.
\autoref{fig:unseen-result} shows the percentage of passing tests for each migration. 
The dashed lines show the median passing tests for each model.
\fo has the highest median passing tests (\UnseenFouroMedianPassedTests), followed by \mn (\UnseenMiniMedianPassedTests).
\lm has most of the migrations (\UnseenLlamaNoPassPercent) fail, resulting in the median passing tests of \UnseenLlamaMedianPassedTests.

\fo has \UnseenFouroAllPass fully correct migrations (i.e., all tests passing after migration) plus two more migrations having 98\% and 99\% tests passing.
\fo has only \UnseenFouroNoPass fully incorrect migrations (no tests passing after migration).
\mn follows \fo with \UnseenMiniAllPass fully correct migrations, and \UnseenMiniNoPass fully incorrect migrations.
\lm has only \UnseenLlamaAllPass fully correct migration but has two migrations having 98\% tests passing.
Overall, the relative performance of the three models follow the same trend as \rqBench and \rqTest, with \fo being the best, followed by \mn and then \lm.

We now compare \rqUnseen results with \rqBench for the \UnseenOldLibPairsCount library pairs that also appear in \migbench.
This helps us understand whether migrations between these library pairs are inherently challenging or if the LLMs did better on potentially familiar code.

\pair{attrs}{dataclasses} and \pair{cryptography}{pycryptodome} migrations illustrate the challenging cases in these four library pairs.
We find that these library pairs show consistent poor performance across all RQs.
In \rqUnseen, the \pair{attrs}{dataclasses} migration results in complete failure across all three models. Similarly in \rqBench, only 2 out of 12 attempts for this library pair by the models are fully correct.
The \pair{cryptography}{pycryptodome} migration completely fails for two models in \rqUnseen, and only 12 out of 45 such instances are correct in \rqBench.

In contrast, \pair{toml}{tomlkit} and \pair{flask}{bottle} represent two consistently easier cases, where the LLMs achieve high or perfect correctness across all RQs.
Except for one \pair{flask}{bottle} migration where the token limit was exceed for \lm in \rqBench, all migrations in these two library pairs are at least partially correct.

Overall, this suggests that the difficulty of a migration is primarily determined by the inherent complexity of the library pairs, not by whether the migration was seen during training. Specifically, library pairs that were hard for the models in \rqUnseen were also hard in \rqBench and \rqTest, while easier pairs consistently achieved better results.





\begin{findingenv}{\rqUnseen}{finding:unseen-correctness}
Out of \UnseenReposCount unseen migrations, the LLMs fully migrated \UnseenLlamaAllPass-\UnseenFouroAllPass migrations, with \fo being the highest.
The performance patterns are consistent with those observed in \rqBench. 
\end{findingenv}

    \section{Discussion}
\label{sec:discussion}

\subsection{Are LLMs Suitable For Library Migration?}

\subsubsection{The Good}

We start with the promising side of using LLMs for library migration.
In addition to the high overall correctness of code changes and migrations in RQ1,
we find that the LLMs can migrate \ccs that were traditionally known as difficult to migrate~\cite{xu2019meditor}. We also find that LLMs can perform migrations on code they have not seen before.
We discuss some examples to demonstrate these strengths.

\paragraph{Different API styles}
The library \lib{argparse} \cite{libargparse} provides API functions for parsing command-line arguments, while \lib{click} \cite{libclick} provides decorators.
Despite the completely different API styles, the three LLMs correctly migrate \CCArgparseClickCorrectPercent of \ccs between this library pair.



\begin{figure}[t]
    \centering
    \footnotesize
    \includegraphics[width=.33\textwidth,trim={0 0 21cm 0},clip]{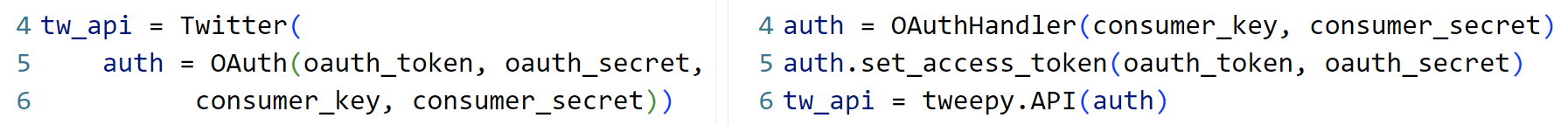} \\
    (a) Before migration\\
    ~\\
    \includegraphics[width=.38\textwidth,trim={18cm 0 0 0},clip]{img/example-mm.png} \\
    (b) After LLM migration
    
    \caption{\lm correctly migrating \mm \cc}
    \label{fig:example-mm}
    \vspace{-0.4cm}
\end{figure}

\paragraph{Higher cardinality migrations}
As shown in RQ1, \lm, \mn, and \fo successfully migrated \LlamaCCBMHigherCardinalityCorrectPercent, \MiniCCBMHigherCardinalityCorrectPercent, and \FouroCCBMHigherCardinalityCorrectPercent of the higher cardinality \ccs, respectively, which the literature always viewed as complex migrations \cite{xu2019meditor,alrubaye2020learning}.
\autoref{fig:example-mm} shows an example of a \mm code change in a \pair{twitter}{tweepy} migration done by \lm.
In addition to correctly replacing the functions \texttt{Twitter} and \texttt{OAuth} with \texttt{OAuthHandler}, \texttt{set\_access\_token}, and \texttt{API} with correct arguments,
\lm also splits the code into multiple statements, making the code more readable.

\paragraph{Inferring changes outside of the source and target APIs}
In an \pair{eventlet}{gevent} migration, \fo replaced a call \texttt{SocketIO()} with \texttt{SocketIO(async\_mode='gevent')}. 
This is interesting because the \texttt{SocketIO} API is from the \lib{Flask-SocketIO} library, which is neither the source nor the target.
The \texttt{SocketIO} function works with several libraries, \lib{eventlet} being the default \cite{flasksocketio}.
If \fo did not set the argument to \texttt{gevent}, the code would break after the migration.


\paragraph{Unseen migrations} When using LLMs for library migration, we do rely on their knowledge of the APIs. However, we also need to ensure they can apply this knowledge to unseen code and are not just reciting the target code.
The results of RQ3 suggest that this is indeed the case, giving us confidence that an LLM-based migration tool can be useful in practice.
RQ3 even demonstrates that LLMs can handle large migrations they have not seen before. 
Specifically, the \pair{attrs}{dataclasses} migration in \rqUnseen is the largest one, with 5 files and 210 source API usages to migrate,
where \fo changed 615 lines of code to perform this migration.
\autoref{fig:unseen-result} shows that all three LLMs failed all tests for this migration.
However, upon further investigation, we find that \fo was able to migrate most of the code correctly, but the tests failed because the original code had a default field defined before a non-default field in a class, which is not allowed in the target library.
We manually fix this by changing 2 lines of code, resulting in 53\% tests passing.
This suggests that, in the worst case, LLMs can provide a good starting point even for large (unseen) migrations that may be more challenging.

\subsubsection{The Bad}

We also analyze cases where the LLMs did not correctly migrate the code and present notable observations.

\paragraph{Failing to handle argument transformation}
The \texttt{wait\_fixed} argument in \texttt{@retry()} from the library \lib{retrying} \cite{libretrying} expects the time in milliseconds, while the target library \textit{tenacity} \cite{libtenacity} expects it in seconds. Therefore, it requires transformation, for example, from \texttt{@retry(wait\_fixed=2000)} to \texttt{@retry(wait=wait\_fixed(2))}.
Out of 28 instances of this change, \fo handled three instances correctly, and the other two models each handled just one instance correctly.
This suggests that while the LLMs can map parameters, differences in expected format or units are harder to manage.

\paragraph{Replacing exception type}
We find cases where the LLM replaces generic exception types with specific ones. \eg \texttt{Exception} with \texttt{ValueError}, which can potentially fail to catch the same exceptions as the original code.
This occurs in both migration-related and non-migration-related changes.
While using more specific exceptions is considered good practice~\cite{cacho2014does}, it does change the program behavior.
Since modern development tools are often equipped to suggest better exception handling, it may be safer for a migration tool to avoid changing existing exception types.

\subsubsection{The Ugly} We also observe the following problematic LLM behavior.

\paragraph{Extra Changes} LLMs are known for not always following the instructions provided in the prompt \cite{ye2023cognitive}. 
Despite our prompt explicitly asking the LLMs to avoid any non-migration changes, we find cases where an LLM, particularly \lm, completely removes parts of the code that do not include any source APIs.
Silently deleting code can be dangerous for the overall functionality of the system, and led to test failures in RQ2 for many cases.

The LLMs also sometimes refactor code that is not related to the migration.
While these improvements can be beneficial, they are unrelated to the migration task at hand.
Future work could explore the use of an interactive IDE plugin that confirms the code changes before applying them, potentially allowing the developer to directly edit the changes.
While less ``dangerous'', we also find that the LLMs remove code comments and docstrings from the source code, even though the comments remain relevant after migration.

\paragraph{Large file handling}
We observe that large files pose a challenge to \lm, given its smaller context size.
Additionally, we observe that even the other two models often migrate an initial part of a large file but then incorrectly state that the remaining part does not require any changes.

\subsection{Differences in Evaluation Setup}
\label{sec:discussion:discrepancies}
The test-based evaluation in \rqTest uses a subset of \rqBench migrations.
Ideally, the same migration in these two RQs should yield the same results.
However, in practice, we observe some discrepancies, which we investigate and explain below.

\subsubsection{Library version compatibility}
In \rqBench, when we manually match changes, we mark a \cc to be correct based on the documentation of latest version of the library.
Since we do not ask for a specific version in the prompt, we notice that the LLMs usually pick APIs from the latest version of the target library, so the LLM's result and manual evaluation align.
However, in \rqTest, we install the versions applicable at the time of migration (Section \ref{sec:unittest-approach}).
This version discrepency often led to run-time errors when running tests on the LLM's code, resulting in the migration marked as incorrect in \rqTest.
\subsubsection{Function parameter shadowing import name}

\begin{figure}[t]
    \centering
    \footnotesize    
    \begin{subfigure}[t]{0.49\columnwidth}
        \centering
        \includegraphics[width=.77\linewidth]{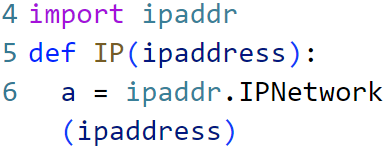}
        \caption{Before migration}
    \end{subfigure}
    \begin{subfigure}[t]{0.49\columnwidth}
        \centering
        \includegraphics[width=.8\linewidth]{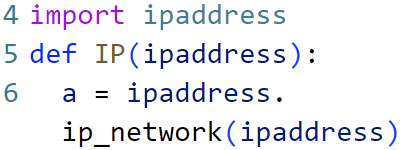}
        \caption{After \mn's migration}
    \end{subfigure}
    \caption{Example of parameter shadowing import name.}
    \label{fig:param-shadow-import-name}
    \vspace{-0.4cm}
\end{figure}

In \autoref{fig:param-shadow-import-name}, \mn correctly replaces the import \texttt{ipaddr} with import \texttt{ipaddress} and the API \texttt{IPNetwork} with \texttt{ip\_network}; therefore, in \rqBench, we count this as a correct migration.
However, notice that the parameter in the function \texttt{IP} is named \texttt{ipaddress}, the same as import name,
which causes the parameter to shadow the import name and leads to test failures.
The developer and \fo both handled this by changing the parameter name (not shown in the Figure), therefore the tests did not fail for them.

\subsubsection{Developer doing non-refactoring changes}
Recall that in RQ1, we find cases where the developer makes non-refactoring changes in the same commit as the migration while the LLM only does the migration it is expected to do.
We count this as a correct migration in RQ1. However, given the changed behavior (often reflected in the tests), the tests fail in \rqTest.

\subsection{Understanding the cost of LLM-based migration}
The \MigExperimentCount migrations we use have a total of 2,584K lines of code (KLOC) for all 10 runs per models. 
The cost of running our experiments was approximately \$\MiniTotalCost for \mn and  \$\FouroTotalCost for \fo, which equates to \$\MiniCostPerKLOC and \$\FouroCostPerKLOC per KLOC, respectively.
\lm is free if self-hosted, though we used a pro subscription of \textit{Hugging Face} (US\$9/month) to avoid installation overheads.
While library migration is an infrequent task in a project's lifetime, it requires significant developer effort when it happens.
The cost of a developer's time can easily exceed the cost of using an LLM, even \fo, which can justify the cost of LLM-based tooling.

    \section{Related work}
\label{sec:related-work}
\subsection{Traditional library migration techniques}
The majority of automated techniques for library migration focus only on identifying API mapping. 
Some techniques identify analogous APIs by analyzing \textit{diffs} in migration commits using static analysis \cite{teyton2013automatic, alrubaye2018automating, alrubaye2019use} and machine learning \cite{alrubaye2020learning}.
Other efforts overcome the need for real migrations by using unsupervised learning \cite{miningAnalogicalAPIs} and natural language processing (NLP) of documentation \cite{ni2021soar}.
DeepMig \cite{deepmig} uses a transformer-based architecture to recommend alternative library and a migration plan, but does not perform the transformation.

SOAR \cite{ni2021soar} applies migration to the client code using program synthesis guided by unit tests.
SOAR is evaluated on two pairs of deep learning libraries, focusing on the migration of neural network models.
The nature of these APIs allows SOAR to check the program state after each API call, which is not the case for most libraries.
Additionally, given the limited number of libraries, their technique relies on the library-specific error message format to further guide the synthesis process, requiring extensive changes and technique adaptations for running SOAR on other libraries.
In contrast, due to the diversity of the libraries we evaluate on, we use the unit tests existing in the repositories to evaluate the correctness of the migrated code.
SOAR supports \om \ccs, but only for a small set of pre-determined APIs.
SOAR times out for 20\% of the evaluated migrations, even for small code snippets (the largest being 183 lines of code).
Our evaluation suggests that LLMs may be able to handle higher-order transformations out of the box for larger code snippets.

\subsection{LLM-based library migration}
Almeida \etal \cite{almeida2024automatic} evaluate the performance of ChatGPT for \pair{SQAlchemy 1} {SQLAlchemy 2} migrations.
Nikolov \etal \cite{nikolov2025google} describe Google's experiences in using LLMs for several internal code transformation tasks, including \pair{JUnit 3}{JUnit 4} and \pair{Joda Time}{Java Time} migrations.
They find that using LLMs reduced migration time by at least 50\% compared to manual migration.
Zhou \etal \cite{hapim} proposed HaPiM, which first trains a machine learning model capable of API mapping,
and then uses the API mappings to guide an LLM in code transformation.
Their approach outperforms MigrationMapper \cite{alrubaye2019use} and GPT-3.5 Turbo on the BLEU \cite{bleu} and CodeBLEU \cite{codebleu} metrics evaluated on 5 Java library pairs.

The above studies show that LLMs can be effective for library migration.
These studies, however, have several limitations.
First, Almeida \etal \cite{almeida2024automatic} and Google's \pair{JUnit 3}{JUnit 4} migrations \cite{nikolov2025google} explore LLMs capability of migrating from one version of a library to another, while we focus on migrating between different libraries.
Migrating versions is different than migrating libraries, because the former can leverage the evolution or release history of the libraries to identify API changes.
Second, all three evaluations are based on very limited number of libraries, maximum 5. 
Google's \pair{Joda Time}{Java Time} migrations \cite{nikolov2025google} also provide library specific instructions. Therefore, there is a lack of generalizability.
Our evaluation, on the other hand, covers \LibPairAllCount Python library pairs from \DomainAllCount application domains.
HaPiM \cite{hapim} also requires first to train a machine learning model to identify API mappings using a set of existing migrations, which is a limitation noted in previous studies \cite{ni2021soar,miningAnalogicalAPIs}.
Finally, none of the above studies analyze the types of \ccs that can be handled.



    \section{Threats to Validity}
\label{sec:threats}
\subsection{Internal validity}
Our benchmark-based evaluation in \rqBench relies on \migbench labeled data, which may contain errors.
The \migbench authors, however, manually vetted each \cc to ensure correctness.
During our manual evaluation, we only found a couple of incorrectly labeled changes, which were eventually corrected in \migbench.

Our evaluation involves two manual steps: validating correctness of \ccs and labeling a non-migration related change as refactoring or non-refactoring.
Manual activities are inherently subjective and may contain errors.
We minimize this by having two authors independently perform the manual activities and resolving disagreements through discussion.

\subsection{Construct validity}

In \rqBench, we analyze the \cc categories based on \changeDev, not on \changeLLM.
The categories can differ if the LLM produces a different \cc than the developer.
Categorizing the LLM changes would require additional manual effort following the previous work by Islam et al.~\cite{pymigtax}, which is beyond the scope of this work.
That said, our observation is that the code/API differences in most cases are minor, and would not yield in a different \taxonomy category.

Tests passing may not always guarantee migration correctness (e.g., the tests do not cover certain edge cases).
We mitigate this, to the extent possible, by selecting repositories with high test coverage in RQ3 and ensuring that the tests cover the source API usages.

\subsection{External validity}
We use three off the shelf LLMs.
While using an LLM specifically tuned for the library migration task could potentially yield better results, the goal of this work is not to develop a migration tool, but to understand the challenges and opportunities of using LLMs for library migration.
Our results may be viewed as a lower bound of the performance of potentially more specialized models.

Our results may not generalize beyond Python.
We choose Python, because it is a popular language, and because we did not find a dataset that has the same detailed characterization of \codechanges for other languages. 

\migbench may not be representative of all Python libraries.
However, the dataset contains a diverse set of \LibPairAllCount library pairs from \DomainAllCount different application domains.
On the other hand, these are all existing libraries where the LLMs are likely familiar with their APIs and API usage.
This familiarity with APIs and API usage is precisely our hope when using LLMs for migration.
However, being familiar with the APIs alone does not always guarantee being able to map them to each other or apply the right transformations on all code bases, which is why our evaluation is necessary.
For newer libraries, additional information, e.g., API documentation, may be needed to guide the LLM in the migration process.
This can be an interesting avenue for future work.
    \section{Conclusion}
\label{sec:conclusion}

This paper presented an empirical study of using LLMs for library migration in Python.
Specifically, we use \llama, \mini, and \fouro to migrate \MigExperimentCount migration commits across \RepoExperimentCount client repositories.
We compared the LLM changes to the developer changes recorded in \migbench for these migrations.
We find that \fo performs the best, and correctly migrates \FouroCCBMTotalCorrectPercent of the individual code changes.
At the migration level, it perfectly migrates \FouroMigBMCorrectPercent of the migrations, and \FouroMigBMAtLeastOneCorrectPercent of them have at least one correctly migrated \cc.
We also run unit tests for a subset of the migrations, and find that \FouroMigTestCorrectPercent of the migrations by \fo have the same sets of tests passing in the developer's migration and the LLM's migration.
A much lower cost LLM \mn and the free \lm also perform relatively well, with both correctly migrating \MiniCCBMTotalCorrectPercent of \ccs.
We also find that the LLMs can correctly perform unseen migrations.
Overall, our results suggest that using LLMs for library migration is promising, and we discuss the opportunities for further work in this area.

    \printbibliography 

\end{document}